\documentclass[twocolumn]{aastex631} 
\usepackage{subcaption}
\usepackage{amsmath}

\begin{document}

\title{
A Dual-Resolution Prescription in the $S_N$ Method for Boltzmann Neutrino Transport I: Proof of Principle and the Resolution of Collision Term
}
\author[0009-0009-8598-7361]{Akira Ito}
\affiliation{Graduate School of Advanced Science and Engineering, Waseda University, 3-4-1 Okubo, Shinjuku, Tokyo 169-8555, Japan}

\author[0000-0002-9234-813X]{Ryuichiro Akaho}
\affiliation{Faculty of Science and Engineering, Waseda University, 3-4-1 Okubo, Shinjuku, Tokyo 169-8555, Japan}

\author[0000-0002-7205-6367]{Hiroki Nagakura}
\affiliation{Division of Science, National Astronomical Observatory of Japan, 2-21-1 Osawa, Mitaka, Tokyo 181-8588, Japan}

\author[0000-0002-2166-5605]{Shoichi Yamada}
\affiliation{Department of Physics, School of Advanced Science and Engineering, Waseda University, 3-4-1 Okubo, Shinjuku, Tokyo 169-8555, Japan}

\begin{abstract}
We propose a dual-resolution prescription meant for the Boltzmann neutrino transport, in which the advection and collision terms are calculated with different angular resolutions in momentum space.
The purpose is to address the issue of the low resolution that afflicts the $S_N$ method in the multi-dimensional neutrino transport simulations for core-collapse supernovae.
We handle with a high resolution the advection term alone, assuming that the collision term does not require such high resolutions.
To confirm this surmise as well as our new conversion scheme, from low- to high-angular resolutions and vice versa, we run a couple of experimental one-zone (in space) simulations.
Neutrino scatterings on nucleons are considered with small recoils fully taken into account whereas the advection term is replaced by the angle- and energy-dependent source terms that are designed to mimic the results of a Boltzmann simulation, inducing the anisotropy in momentum space.
For the conversion from a low-resolution distribution function to a high-resolution one, we employ a polynomial interpolations in the zenith and azimuth directions separately with the number conservation and continuity (and periodicity only in the azimuth direction).
We find that this dual-resolution scheme works well and that the current angular resolution employed in the canonical supernova simulations with our Boltzmann solver or a bit better in the $\phi_\nu$ direction will be sufficient for the collision terms if they are coupled with the advection terms calculated with a high-angular resolution via this prescription.

\end{abstract}
\section{Introduction} \label{sec:intro}
Core-collapse supernovae (CCSNe) are explosive events that finish the lives of massive stars with $ M_{\mathrm{ZAMS}} \gtrsim 8 M_\odot$.
Since they involve various physical processes coupled with one another nonlinearly, revealing the explosion mechanism of CCSNe is a challenging problem across astro-, nuclear and particle physics.
It requires solving hydrodynamics and neutrino transport (and gravity) quantitatively and simultaneously.
Although we have witnessed remarkable progresses in the theoretical understanding thanks to the growth of computational resources and the advances in numerical methods over several decades, and multi-dimensional simulations produce explosions rather commonly these days \citep{Burrows2021,Burrows2024,Janka2024}, the details of the explosion mechanism are still not fully uncovered \citep{janka2016,mezzacappa2023,Yamada2024,boccioli2024}.

It is a well-known fact that the neutrino heating is crucial for explosion.
Since neutrinos are neither in thermal nor in chemical equilibrium in general, their transport is described by a kinetic equation, for example, the Boltzmann equation if the quantum kinetics is ignored.
They may be solved either by deterministic methods like the $S_N$ method \citep{Mezzacappa1993,liebendorferProbingGravitationalWell2001,Sumiyoshi2005,sumiyoshiNEUTRINOTRANSFERTHREE2012,nagakuraThreedimensionalBoltzmannHydroCode2014,akaho2021} or by probabilistic methods such as the Monte Carlo method \citep{Fleck_Cummings1971,Fleck1984,Keil_Raffelt_Janka2003,Abdikamalov2012,richersDetailedComparisonMultiDimensional2017,katoNeutrinoTransportMonte2020a,kato2021}.
As an aside, they may also be solved in the context of the neutron star merger simulations \citep{foucart2018,miller2019,foucart2021,kawaguchi2023,Nils2024,kawaguchi2025_longterm,kawaguchi2025_bhtorus}.
Either way, numerically solving the classical Boltzmann equation, not to mention the quantum kinetic equation, requires considerable computational resources \citep{Mezzacappa1993,Mezzacappa1999,Liebendorfer2002,Sumiyoshi2005,sumiyoshiNEUTRINOTRANSFERTHREE2012,Peres2014,Chan2020}.
In order to reduce the computational costs, various approximate methods have been conceived$\colon$the flux-limited diffusion approximation (FLD)\citep{Arnett1977,Janka1992}, the isotropic diffusion source approximation (IDSA)\citep{Liebend_Whitehouse_Fischer2009,Takiwaki2012}, the ray-by-ray (plus) approximation (RbR+)\citep{Rampp2002,burasTwodimensionalHydrodynamicCorecollapse2006}, and the truncated moment methods with prescribed closure relations \citep{Thorne1981,Shibata2011}. In exchange for the reduction in the numerical cost, these methods shred some information on the neutrino distribution in momentum space.

In principle, solving the Boltzmann equation faithfully is desirable.
However, a resolution issue arises with the $S_N$ method, from its high numerical cost mentioned above, as observed in our own simulations.
In particular, the momentum space is covered with a rather small number of angular mesh points (10 for $\theta_{\nu}$, the zenith-angle measured from the local radial direction, and 6 for $\phi_{\nu}$, the azimuth-angle, in our typical computations) and the numerical solution tends to deviate from the true solution in the optically thin region, where the neutrino distribution function becomes remarkably forward-peaked \citep{richersDetailedComparisonMultiDimensional2017}.
It is true that the neutrino transfer simulation with the Boltzmann solver is expensive and time-consuming but it is still important, particularly for gauging possible systematic issues with the approximate methods in realistic dynamical settings.
It is hence crucial to address the resolution issue.

The key observation here is that the numerical error is mainly caused by the advection term; the collision term suffers less from the angular resolution, since it is insignificant by definition in the optically thin region, where the neutrino distribution is substantially anisotropic.
On the other hand, the collision term is numerically stiff in the optically thick region.
Hence it is a good idea to take advantage of these differences in the characteristics of the advection and collision terms and adopt different resolutions for them.
Our strategy is hence to increase the angular resolution only of the advection term while keeping the resolution of the collision term low at the current level.

As a proof of principle, we run experimental one-zone (in space) simulations in this paper.
We confirm that the current level of angular resolution is indeed adequate for the collision terms, testing at the same time our new prescription for the transformations between different resolutions, both in time-independent and time-dependent settings. 
In particular, the conversion from a lower- to a higher-resolution is non-trivial.
It is impossible in fact, mathematically speaking.
However, we pursue a method with acceptable errors.
Ideally, we should do this test by running a Boltzmann solver with the advection and collision terms with different angular resolutions fully implemented.
It is admittedly costly, though.
Considering the nature of this study as a proof of principle, we take a simpler approach, in which one-zone calculations are performed by utilizing the source terms that are designed to mimic the spatial advection.
Note that the treatment of the advection term is rather trivial: we have only to increase the number of angular grid points. As we mentioned above, what is truly non-trivial is the transformations of neutrino distribution functions with different angular resolutions.

Sec.~\ref{sec:Boltz} gives the overview of our Boltzmann project.
The numerical methods are described in Sec.~\ref{sec:style}; in particular, the one-zone model we employ in this paper is explained in  Sec.~\ref{sec:validation}.
The results are presented in the following two sections: Sec.~\ref{sec:result} and Sec.~\ref{sec:time_result}.
Finally, in Sec.~\ref{sec:Conclusion}, we summarize our results with discussions.

\section{Numerical methods}

\subsection{Boltzmann Transport} \label{sec:Boltz}

The neutrino transfer may be described by the Boltzmann equation if the quantum effect is ignored:
\begin{equation}\label{eq:boltz}
    p^\mu \frac{\partial f}{\partial x^\mu} - \Gamma^i_{\mu\nu} p^\mu p^\nu \frac{\partial f}{\partial x^i} = - p^\mu u_\mu C[f],
\end{equation}
where $f$ is the neutrino distribution function in phase space, $x^\mu$ is the spacetime coordinates, $p^\mu$ is the four-momentum of neutrino, $\Gamma^i_{\mu\nu}$ is the Christoffel symbol, $u_\mu$ is the four-velocity of matter and $C[f]$ is the collision term; here index $i$ runs from $1$ to $3$ while $\mu$ and $\nu$ run from $0$ to $3$.
The first and second terms on the left hand side of Eq.~\eqref{eq:boltz} describe the advections in space and momentum space, respectively.
The latter is nonvanishing even in the absence of external agents such as magnetic fields if the spacetime is not flat or curvilinear coordinates are employed even in the flat spacetime.
The collision term on the right hand side describes the interactions between neutrinos and matters or among neutrinos themselves and is given as integrals over momentum space at each spatial point.
It is nonlinear in the distribution function if neutrino-pair processes are taken into account or the Pauli-blocking of neutrinos in the final state of scatterings is incorporated.
Hence the Boltzmann equation is a nonlinear integro-differential equation in general.

Over the years we have performed CCSN simulations with our Boltzmann solver that employs the $S_N$ method and is coupled with a hydrodynamics code.
They are mostly 2D simulations under the assumption of axisymmetry of the system \citep{nagakuraSimulationsCorecollapseSupernovae2018,Nagakura2019,Harada_Nagakura_Iwakami_Okawa_Furusawa_Matsufuru_Sumiyoshi_Yamada2019,Harada2020,akaho2021,Harada2022,Iwakami2022,akaho2023}
but 3D simulations are also underway \citep{sumiyoshiNEUTRINOTRANSFERTHREE2012,nagakuraThreedimensionalBoltzmannHydroCode2014,nagakura2017,nagakuraComparingTreatmentsWeak2019,Iwakami2020}.
In the $S_N$ method, the derivatives in the advection terms are replaced with the finite-differences in all coordinate directions in phase space; the integrals in the collision term are approximated by finite sums.
In this process we normally handle the two terms with the same resolution, i.e., on the same numerical mesh.
The Boltzmann equation is six dimensional (3 in space plus 3 in momentum space) and the inversion of the matrix is computationally demanding \citep{Mezzacappa1993,Mezzacappa1999,Liebendorfer2002,Sumiyoshi2005,sumiyoshiNEUTRINOTRANSFERTHREE2012,Peres2014,Chan2020}. 
\cite{richersDetailedComparisonMultiDimensional2017} showed by the comparison with his Monte Carlo simulations that the Eddington tensor may be underestimated by $ \sim 10 \% $ in the optically thin region, where the neutrino distribution becomes highly forward-peaked, and that the angular resolution may need to be $\sim 4$ times higher than the current one in order to get a good convergence of the numerical solution with our Boltzmann solver.

The deviation from the true solution comes mostly from the evaluation of the advection terms with the low angular resolution in momentum space.
Simply increasing the angular resolution is not feasible as mentioned above, though.
Our strategy here is then to increase the angular resolution only for the evaluation of the advection terms.
To employ different resolutions for the advection term and the collision term, we adopt the operator-splitting method, which is commonly adopted in the truncated moment methods \citep{Rampp2002,Skinner2019,Laiu_Endeve2021}.
For example, \cite{Laiu_Endeve2021} employ the DG-IMEX scheme based on the discontinuous Galerkin method and the implicit-explicit time integration.
However, it has not been attempted in the Boltzmann transport so far.
The most crucial ingredient in the use of dual resolutions is the conversion of the distribution functions obtained with the different resolutions.
In this paper we apply different interpolation schemes based on polynomials to the zenith and azimuth directions separately.

This is a pilot study.
A full implementation of the scheme thus confirmed to our Boltzmann solver is beyond its scope and is postponed to the next study.

\subsection{Numerical strategy} \label{sec:style}

In this subsection, we explain more concretely our ideas to treat the advection and collision terms with different resolutions.
As stated before, they consist of two elements: (1) the operator-splitting treatment of the two terms and (2) the conversion of a distribution at one resolution to a distribution at another resolution.
They will be given in Sec.~\ref{sec:OneZone} and Sec.~\ref{sec:remap}, respectively.

In this paper, we focus on the neutrino scatterings on nucleons, the dominant scattering process in the supernova core. In the previous version of our Boltzmann solver, it is assumed to be elastic, i.e., the energy exchange between neutrino and nucleon is vanishing. It has been demonstrated \citep{Raffelt2001,Rampp2002,Thompson2003,Burrows2004,burasTwodimensionalHydrodynamicCorecollapse2006,Marek2009,Hudepohl2010,Pllumbi2015,melsonNeutrinodrivenExplosion202015,Skinner2016,Radice2017,kotakeImpactNeutrinoOpacities2018,suwaKompaneetsEquationNeutrinos2019,Vartanyan2019,burrowsThreedimensionalSupernovaExplosion2019,wangGeneralizedKompaneetsFormalism2020,Bruenn2020,kato2021}, however, that nucleon recoils as well as other corrections of $O(q/M_N)$ have important roles to play for shock revival, where $q$ is the magnitude of momentum transfer and $M_N$ is the nucleon mass. In this paper, we consider the recoil alone for simplicity and neglect the weak magnetism and other form factors in the weak currents as well as the possible contribution from strange quarks, i.e., the isoscalar correction to the axial vector current \citep{albericoStrangenessNucleonNeutrinonucleon2002,leitnerNeutrinoInteractionsNucleons,melsonNeutrinodrivenExplosion202015,hobbsRoleNucleonStrangeness2016}, and the many-body effects \citep{burrowsEffectsCorrelationsNeutrino1998,mornasNeutrinonucleonScatteringRate2001,mornas2003,horowitzNeutrinonucleonScatteringSupernova2017,burrowsCrucialPhysicalDependencies2018} other than the effective nucleon masses. In order to incorporate the small energy exchange by recoil, we deploy a subgrid in energy \citep{burasTwodimensionalHydrodynamicCorecollapse2006}.
We explain in Appendix its implementation rather in detail.

This paper is meant to be a proof of principle.
We hence adopt a toy-model approach instead of running the Boltzmann code with the new method fully implemented:
we conduct spatially one-zone calculations, in which the advection term is replaced with the source term, the details of which will be explained in the last subsection \ref{sec:validation}.
One may wonder if such an approach can really address the problem.
As we mentioned earlier, the evaluation of the advection term with a high-angular resolution is rather trivial and is not an issue.
What is nontrivial is the transformations between the low- and high-resolutions as well as the determination of an acceptable low resolution for evaluation of the collision term.
They can be investigated quantitatively in the toy model considered in this paper as long as the source term is appropriately chosen. This will be explained further in Sec.~\ref{sec:validation}.

\subsubsection{Operator splitting}\label{sec:OneZone}
The operator-splitting treatment of the advection and collision terms in the transport equation is rather common particularly in the truncated moment method \citep{Rampp2002,burasTwodimensionalHydrodynamicCorecollapse2006} and some advanced formulations like the IMEX method have been developed \citep{oconnor2015open,just2015newtrans,kuroda2016newtransGR,Skinner2019,Chu_Endeve2019,Laiu_Endeve2021}.
In this paper, we do not pursue such sophistications, since our purpose is to demonstrate that the separate treatment of the advection and collision terms with different angular resolutions is possible.
We will incorporate one of such advanced methods later in the actual implementation to the Boltzmann solver. We hence give here a brief description of what we do in this paper.

Suppose that we have calculated the distribution function up to the $n$-th time step and advance the calculation one step further to the $(n+1)$-th time step.
We first calculate the contribution from the advection term alone in an explicit manner:
\begin{eqnarray}
    \frac{f^\ast - f^{n}}{\Delta t} = A[f^n],
\end{eqnarray}
where $f^n$ is the distribution function at the $n$-th time step and $f^\ast$ is the distribution function at the intermediate step; $A$ stands for the advection term symbolically, which is replaced in this paper by the source term that will be explained later in Sec.~\ref{sec:validation}.

\begin{figure}
  \begin{minipage}{1\linewidth}
    \centering
    \includegraphics[width=1\textwidth]{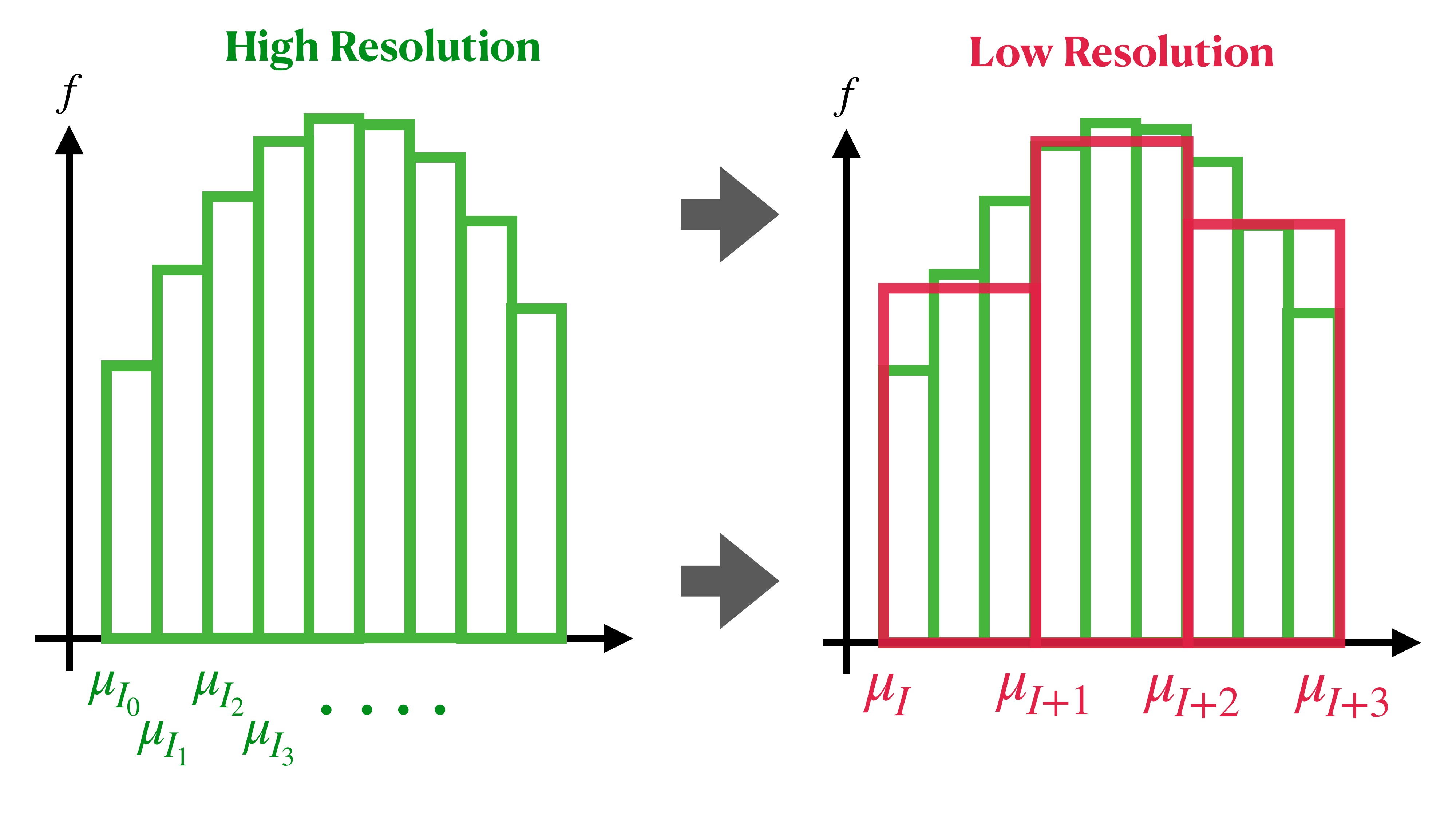}
    \subcaption{}
    \label{fig::ponchi-e_1}
  \end{minipage}\\
  \begin{minipage}{1\linewidth}
    \centering
    \includegraphics[width=1\textwidth]{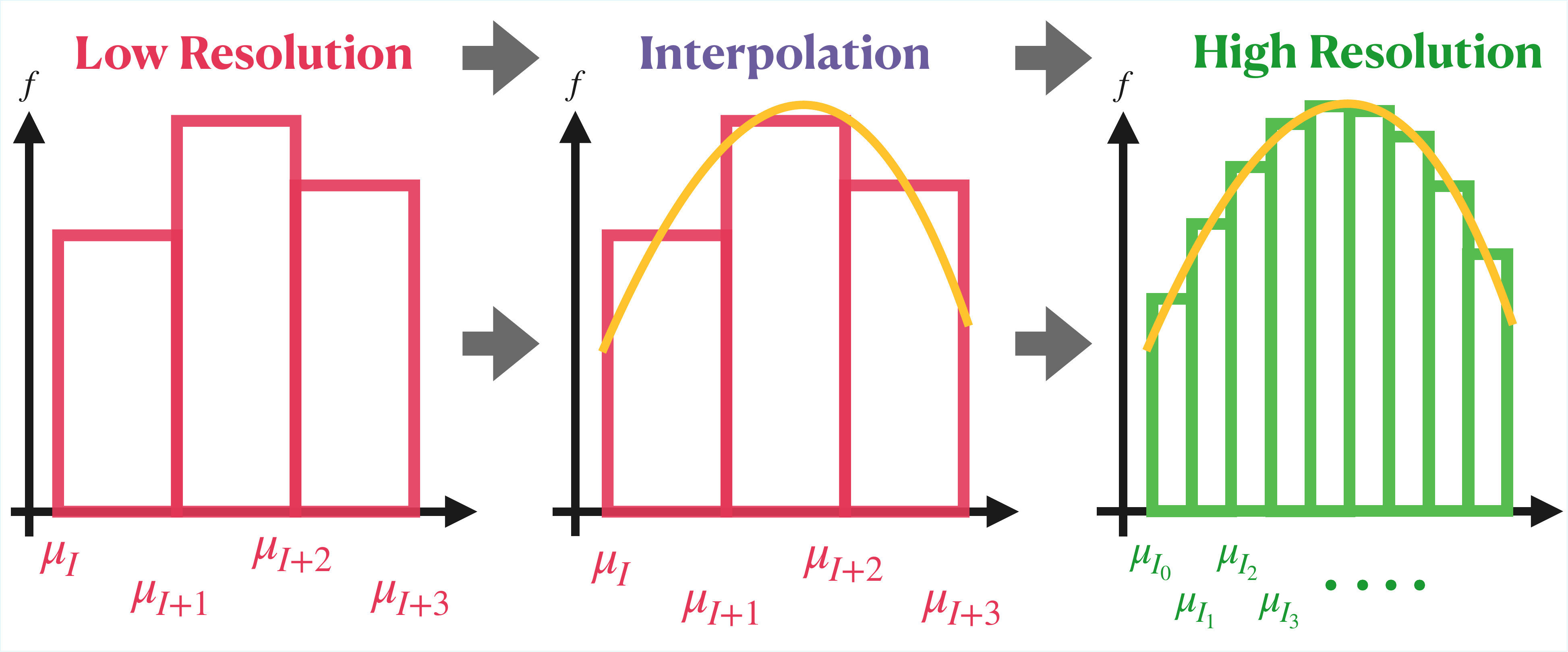}
    \subcaption{}
    \label{fig::ponchi-e_2}
  \end{minipage}
  \caption{Schematic pictures of the resolution conversions. (a) from high- to low-resolutions, (b) from low- to high-resolutions.}
  \label{fig::ponchi}
\end{figure}

Next, we compute the contribution from the collision terms alone.
Before doing so, we need to convert $f^\ast$ obtained with the higher resolution to the distribution $f^\ast_\mathrm{low}$ of lower resolution.
This conversion is shown schematically in Fig.~\ref{fig::ponchi-e_1}.
The concrete method is given in the next subsection.

The collision terms are calculated then at this low resolution, this time in a time-implicit manner as follows:
\begin{eqnarray*}
    \frac{f^{n+1}_{\mathrm{low}} - f^{\ast}_{\mathrm{low}}}{\Delta t} = C[f^{n+1}_\mathrm{low}].
\end{eqnarray*}
The concrete expression of the collision terms is given in Appendix \ref{sec:subgrid}.
This is a system of nonlinear algebraic equations and the linearized equations are solved repeatedly until a certain convergence criterion is satisfied. 

Finally, as shown schematically in Fig.~\ref{fig::ponchi-e_2}, the distribution function at the $(n+1)$-th time step $f^{n+1}_{\mathrm{low}}$ of low resolution is converted back to the higher-resolution distribution function $f^{n+1}$, which closes the entire procedure of the single time step.
Then we move on to the next time step.
The concrete method to convert resolutions will be given in the next subsection.

 \subsubsection{ Conversions of resolutions }\label{sec:remap}
 
 \begin{figure}
  \begin{minipage}{1\linewidth}
    \centering
    \includegraphics[width=1\textwidth]{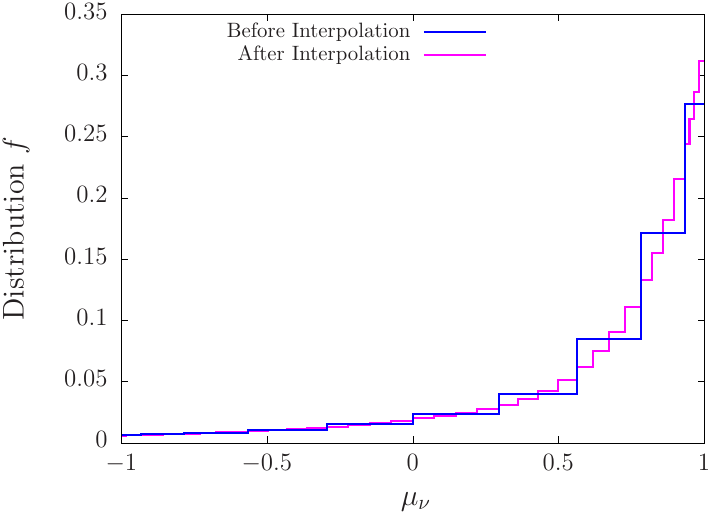}
    \subcaption{}
    \label{fig:interpolate_distribution_in_theta}
  \end{minipage}\\
  \begin{minipage}{1\linewidth}
    \centering
    \includegraphics[width=1\textwidth]{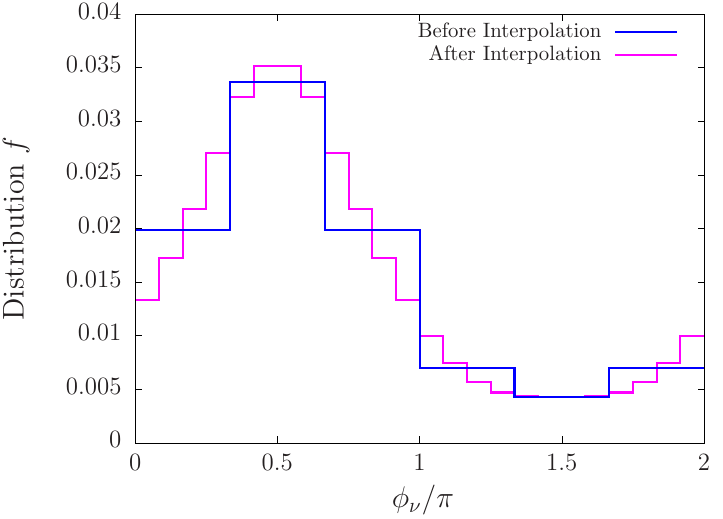}
    \subcaption{}
    \label{fig:interpolate_distribution_in_phi}
  \end{minipage}
  \caption{
  Representative results of interpolations (a) for the $\theta_\nu$ direction and (b) for the $\phi_\nu$ direction.
  In (a) the blue and purple lines correspond to the distributions in $\theta_\nu$ with $N_{\theta_\nu} = 10$ and $N_{\theta_\nu} = 40$, respectively, whereas in (b) they show the distributions in $\phi_\nu$ with $N_{\phi_\nu} = 6$ and $N_{\phi_\nu} = 24$, respectively.
  }
  \label{fig:interpolate_distribution}
\end{figure}
The transformation of a high-resolution distribution to a low-resolution one is straightforward as long as the cell boundaries of the two angular meshes coincide with each other, which we assume here.
As shown in Fig.~\ref{fig::ponchi-e_1}, all we have to do is to combine the neutrino distribution functions for the cells in the finer mesh that are included in a single cell in the coarser mesh.
This is done for the zenith- and azimuth-angles separately.
We assume that the distribution function is constant inside the coarse cell when calculating the collision terms (see Appendix \ref{sec:subgrid}).

The conversion of resolutions in the opposite direction is non-trivial.
It is in fact impossible, rigorously speaking.
The goal is hence to find a reasonable approximation.
Here we try interpolations based on the polynomial fitting to the lower-resolution distribution.

We begin with the treatment of the zenith-angle ($\theta_\nu$) direction.
In the neighborhood of each grid point in the low-resolution mesh, we approximate the distribution function with an $N_{\theta_\nu}^\mathrm{poly}$-th order polynomial of $ \mu_\nu = \cos\theta_\nu$.
It is expressed as
\begin{equation}
    f_{\mathrm{appr}}(\mu_\nu) = \sum_{n=0}^{N_{\theta_\nu}^{\mathrm{poly}}} a_{n} \mu_{\nu}^{n}.
    \label{eq::appr_distribution_theta_nu}
\end{equation}
$N_{\theta_\nu}^{\mathrm{poly}}$, which is a parameter to be determined, should be smaller than $N_{\theta_{\nu}}+2$, the number of cells in the $\theta_{\nu}$ direction in the low-resolution mesh plus the number of boundary conditions.
In the following, we vary both $N_{\theta_{\nu}}^{\mathrm{poly}}$ and $N_{\theta_{\nu}}$ independently under this constraint.

For given $N_{\theta_\nu}^{\mathrm{poly}}$ and $N_{\theta_\nu}$ the coefficients $a_n$ in the polynomial are determined with two steps.
In first step, we determine the values of the distribution function at $ \mu_\nu ( = \cos \theta_\nu ) = \pm 1 $ as follows.
At each grid point of $\theta_\nu$ in the low-resolution mesh, we take an average of the distribution functions over $\phi_\nu$ and build a mean distribution in $\theta_\nu$, $f^{\mathrm{mean}}(\theta_\nu)$.
If the distribution is axisymmetric, this procedure is not necessary.
Then we approximate it with an $(N_{\theta_{\nu}}^{\mathrm{poly}}-2)$-th order polynomial of $\mu_\nu$ in each neighborhood of $\mu_\nu = -1$ and $1$, individually unless $N_{\theta_\nu}^{\mathrm{poly}} = N_{\theta_\nu} + 1$.
In the latter case, the two neighborhood coincide with the entire range $[-1,1]$.
Each polynomial is expressed as
\begin{equation}
    f^{\mathrm{mean}}_{\mathrm{appr}}( \mu_\nu )  = \sum_{n^\prime = 0}^{N_{\theta_{\nu}}^{\mathrm{poly}}-2} a_n^\prime \mu_\nu^n,
\end{equation}
and is determined by the following conditions for the $N_{\theta_\nu}^{\mathrm{poly}}-1$ consecutive cells from the boundary:
\begin{eqnarray}
f^{\mathrm{mean}}(\mu_{i_k}) ( \mu_{I_k} - \mu_{I_{k-1}} ) 
&& = \int^{\mu_{I_k}}_{\mu_{I_{k-1}}} d\mu_\nu f^{\mathrm{mean}}_{\mathrm{appr}}(\mu_\nu) \notag \\
= \sum_{ n = 0 }^{N_{\theta_\nu}^\mathrm{poly}} \frac{a^\prime_{n}}{n+1} && ( (\mu_{I_k})^{n+1} - (\mu_{I_{k-1}})^{n+1} ), \label{eq:theta_poly_mean}
\end{eqnarray}
where $f^\mathrm{mean}$ is the mean distribution function in $\theta_\nu$ and $k$ runs from $0$ to $N_{\theta_{\nu}}^{\mathrm{poly}}-2$, specifying the cells.
Note that we distinguish here the cell-interface values
$\mu_I$ from the cell-center ones $\mu_i$ \citep{sumiyoshiNEUTRINOTRANSFERTHREE2012}.
Equation \eqref{eq:theta_poly_mean} requires that the polynomial should reproduce the neutrino number densities exactly at these cells.
We evaluate the distribution functions so obtained at $\mu_\nu = \pm 1 $.

In the second step, we approximate the low-resolution distribution function $f(\theta_\nu,\phi_\nu)$ at each $\phi_\nu$ separately with a polynomial given in Eq.~\eqref{eq::appr_distribution_theta_nu} again.
Here and in the rest of this paper we assume that the interpolations are done always in the zenith ($\theta_\nu$) direction first.
We determine these polynomials by requiring again that the neutrino number densities should be exactly reproduced for all angular bins in the neighborhood of a grid point of concern.
This time the number of those cells is $N_{\theta_\nu}^{\mathrm{poly}}+1$ if the neighborhood contains no boundary cell.
It is $N_{\theta_\nu}^{\mathrm{poly}}$ if one of the two boundary cells is included and $N_{\theta_\nu}^{\mathrm{poly}}-1$ if both boundary cells are contained.
In the latter two cases, we require further that the polynomial should reproduce the boundary value(s) exactly.
These conditions are expressed as follows: for each cell specified by $i = 1, ..., N_{\theta_{\nu}}$ in the low-resolution mesh, we deploy a polynomial $f_{\mathrm{appr}}$ and require
\begin{eqnarray}
f_\mathrm{low}(\mu_{i_k}) ( \mu_{I_k} - \mu_{I_{k-1}} ) 
&& = \int^{\mu_{I_k}}_{\mu_{I_{k-1}}} d\mu_\nu f_{\mathrm{appr}}(\mu_\nu) \notag \\
= \sum_{ n = 0 }^{N_{\theta_\nu}^{\mathrm{poly}^\ast}} \frac{a_{n}}{n+1} && ( (\mu_{I_k})^{n+1} - (\mu_{I_{k-1}})^{n+1} ),
\end{eqnarray}
\begin{eqnarray}
f_{\mathrm{low}} (\mu_\nu = 1) && = f^{\mathrm{mean}} (\mu_\nu = 1), \notag \\
f_{\mathrm{low}} (\mu_\nu = - 1) && = f^{\mathrm{mean}} (\mu_\nu = - 1), \label{eq:theta_poly_temp}
\end{eqnarray}
where $k$ runs from $0$ to $N_{\theta_{\nu}}^{\mathrm{poly}}$ and specifies the cells included in the neighborhood of cell $i$; $N_{\theta_\nu}^{\mathrm{poly}^\ast}$ is the actual order of the polynomial determined as mentioned above.

The neighborhood of a cell in the low-resolution mesh is chosen in principle so that it should be centered at the cell.
In practice, if the number of cells included in this neighborhood is odd, it is exactly centered at the target cell, whereas if it is even, the neighborhood is extended to the right, that is, to the side of larger $\mu_\nu$.
When it comes close to the mesh boundaries, however, this choice cannot be done.
In that case, we extend the neighborhood in the direction opposite to the boundary: at $i=1$, for example, the neighborhood is completely one-sided.
If the distribution function is axisymmetric, having no $\phi_\nu$-dependence, we skip the second step and $N_{\theta_\nu}^{\mathrm{poly}^\ast} = N_{\theta_\nu}^{\mathrm{poly}}$ simply.

Once the polynomial is determined for each cell in the low-resolution mesh, the distribution functions for the cells in the high-resolution mesh inside that particular low-resolution cell are obtained by integrating the polynomial in those fine cells in the high-resolution mesh as follows:
\begin{eqnarray}
f(\mu_{i_l}) = \frac{1}{( \mu_{I_l} - \mu_{I_{l-1}} ) } \int^{\mu_{I_l}}_{\mu_{I_{l-1}}} d\mu_\nu \sum_{n=0}^{N_{\theta_\nu}^{\mathrm{poly}^\ast}} a_{n} \mu_\nu^{n} \notag \\ 
= \frac{1}{( \mu_{I_l} - \mu_{I_{l-1}} ) } \sum_{ n = 0 }^{N_{\theta_\nu}^{\mathrm{poly}^\ast}} \frac{a_{n}}{n+1} ( (\mu_{I_l})^{n+1} - (\mu_{I_{l-1}})^{n+1} ), \label{eq:theta_poly} 
\end{eqnarray}
where $l$ runs from $1$ to $N^{\mathrm{src}}_{\theta_{\nu}}$, the number of the fine cells inside the single coarse cell.

A representative result in the $\theta_\nu$ interpolation is exhibited in Fig.~\ref{fig:interpolate_distribution_in_theta}, where the distribution function obtained with a lower-angular resolution of $ N_{\theta_\nu} = 10 $ is compared with the reconstructed distribution function with a higher-angular resolution of $ N_{\theta_\nu} = 40 $.
We assume that the distributions are both constant in their own mesh cells.
One finds that the reconstructed distribution function reproduces the original trend without generating artificial features but expresses the forward peak more sharply as expected.
We hence judge that the polynomial interpolation works properly.

For $\phi_\nu$, the same strategy does not work well, since a high-order polynomial introduces undesirable oscillatory behavior in the interpolant.
This is also the case for an interpolant based on the Fourier expansion, which we will not show here. 
This happens because the $\phi_{\nu}$ dependence of the neutrino angular distribution is in general weaker than its $\mu_{\nu}$ dependence.
We find that the following method works rather well  in such situations.
In each cell of the low-resolution mesh we deploy a polynomial of the second order in $\phi_\nu$ as follows: 
\begin{eqnarray}
    f^i_{\mathrm{appr}}(\phi_{\nu}) = b^i_2 \phi_\nu^2 + b^i_1 \phi_\nu + b^i_0,
\end{eqnarray}
where $i ( = 1,\dots,N_{\phi_\nu})$ specifies the cell in the low-resolution mesh.

The $N_{\phi_{\nu}} \times 3$ coefficients, $b_{n}^i$, are determined by requiring that the neutrino number densities should be reproduced exactly in each cell again:
\begin{eqnarray}
f_{\mathrm{low}}(\phi_{\nu,i}) ( \phi_{\nu,I} - \phi_{\nu,I-1}) = && \int_{\phi_{\nu,I-1}}^{\phi_{\nu,I}} d \phi_{\nu} f^i_{\mathrm{appr}}(\phi_\nu) \notag \\
= \sum_{ n = 0 }^{2}  \frac{b_{n}^{i}}{n+1} ( (\phi_{\nu,I})^{n+1} && - (\phi_{\nu,I-1})^{n+1} ), \label{eq:low_range}
\end{eqnarray}
where $I$ is the label for the cell interfaces; both $i$ and $I$ run from $1$ to $N_{\phi_\nu}$.
This time we require further that the quadratic functions deployed in each cell should be joined continuously and smoothly at all cell interfaces including at $\phi_{\nu} = 2\pi$, which ensures the periodicity in the azimuthal direction of the angular distribution function as a whole. These conditions are written as
\begin{eqnarray}
    \sum_{n=0}^{2} b_{n}^i \phi_{\nu,I}^n &=& \sum_{n=0}^{2} b_{n}^{i+1} \phi_{\nu,I}^n, \label{eq::periodic2} \\
    \sum_{n=1}^{2} n b_{n}^i \phi_{\nu,I}^{n-1} &=& \sum_{n=1}^{2} n b_{n}^{i+1} \phi_{\nu,I}^{n-1}, \label{eq::periodic4} 
\end{eqnarray}
where $I(=i)$ runs from $1$ to $N_{\phi_\nu}$ with $b_n^{N_{\phi_{\nu}}+1} = b_n^{1}$. 

Just as in the $\theta_\nu$ direction, once the interpolant for $\phi_\nu$ is determined for each cell in the low-resolution mesh, the values of the distribution functions assigned to the individual cells in the high-resolution mesh are obtained by integrating the quadratic polynomial in those fine cells as follows:
\begin{eqnarray}
f(\phi_{\nu,i}) && = \frac{1}{( \phi_{\nu,I} - \phi_{\nu,I-1} ) } \int^{\phi_{\nu,I}}_{\phi_{\nu,I-1}} d\phi_\nu \sum_{n=0}^{2} b_{n}^i \phi^{n}_{\nu} \notag \\ 
= && \displaystyle\frac{ \displaystyle\sum_{ n = 0 }^{2} \displaystyle\frac{b_{n}^i}{n+1} ( (\phi_{\nu,I})^{n+1} - (\phi_{\nu,I-1})^{n+1} ) }{( \phi_{\nu,I} - \phi_{\nu,I-1} ) }, \label{eq:phi_poly_0pi} 
\end{eqnarray}
where $i$ and $I$ denote the center and interface, respectively, of the high-resolution mesh this time.
As we mentioned earlier, the $\phi_\nu$ interpolation is conducted after the $\theta_\nu$ interpolation is done.
The former interpolation is performed for the values of the distribution functions evaluated on the $\theta_\nu$ grid points in the high-resolution $\theta_\nu$-mesh for each $\phi_\nu$ grid point in the low-resolution $\phi_\nu$-mesh.

A representative result in the $\phi_\nu$ interpolation is presented in Fig.~\ref{fig:interpolate_distribution_in_phi}, where the distribution function obtained with a lower-angular resolution of $N_{\phi_\nu} = 6 $ is compared with the reconstructed distribution function with a higher-angular resolution of $N_{\phi_\nu} = 24 $.
Again we assume that the distribution functions are constant in their own mesh cells. 
It is also apparent that the result is qualitatively satisfactory with no undesirable artifacts.

\subsubsection{One-zone model with Source Terms} \label{sec:validation}

We next move on to the treatment of the advection terms in this paper.
Since our main purpose here is to study the angular-resolution issues, particularly the appropriate resolution for the collision term and the conversion of resolutions, the advection terms are not treated squarely but are modeled as explained shortly.

The spatially finite-differenced form of the Boltzmann equation may be expressed as 
\begin{equation}
    \frac{1}{c} \frac{\partial f_i }{\partial t } + A_{i}[f_{i-1},f_{i},f_{i+1}] = C_{i} [f_i] ,
    \label{eq::boltzmann_represent}
\end{equation}
where $f_i$ is the neutrino distribution at the spatial grid point specified with $i$; all other labels, e.g., the index for neutrino energy, angular in both space and momentum space, are omitted for simplicity.
The term $A_{i}$ in the above equation represents the advection terms at the same spatial point, whereas the term denoted by $C_{i}$ corresponds to the collision terms there. 
The arguments of these terms in Eq.~\eqref{eq::boltzmann_represent} reflect their natures:
the advection terms are non-local in the sense that they depend not only on the neutrino distribution on the same spatial grid point but also on those at nearby grid points;
the collision term, on the other hand, is local, being determined solely by the distribution on the grid point of concern.
As a result, the advection terms couple the equations at different spatial grid points, making it much more involved (and difficult) to solve the equations.
If the advection term were local, the calculation would be considerably easier.
That is what we want to do in this paper and what we will explain shortly below.

It is probably suggestive to recast Eq.~\eqref{eq::boltzmann_represent} as follows: 
\begin{equation}
    \frac{1}{c} \frac{\partial f_i }{\partial t } = C_{i} [f_i] + S_i,
    \label{eq::Boltz_with_source}
\end{equation}
where the second term on the right hand side, $S_i$, is nothing but the advection terms with the sign reversed: $S_i = - A_{i}[f_{i-1},f_{i},f_{i-1}] $.
In this paper, we replace it with the same function but evaluated for known distribution functions, $f_B$, that are obtained in one of our Boltzmann simulations of CCSNe.
By this replacement, $S_i$ is a known function now.
Then the equations for different spatial grid points are all decoupled and can be solved independently of each other.
As mentioned earlier, this makes it very much easier to solve them numerically and enables us to focus on the collision terms alone and deploy a (substantially) larger number of the angular grid points for the resolution study.
Moreover, the resultant solution should be equal to $f_B$ by construction (see the explanation given below), the fact we will use to evaluate the error produced by a limited angular resolution in momentum space.
Note that we may incorporate into $A_i$ also the collision terms other than the nucleon scatterings, the feature we actually employ in the following.
Since Eq.~\eqref{eq::Boltz_with_source} can be solved pointwise, we call the above treatment a one-zone model and refer to $S_i$ as its source term.

In order to validate the dual-resolution strategy proposed in this paper, we perform two types of test calculations using the one-zone model. We refer to them as the steady-state test and the time-evolution test.
In the former test, we give a time-independent source term and integrate Eq.~\eqref{eq::Boltz_with_source} numerically to obtain the steady solutions for different angular resolutions in momentum space.
We compare the asymptotic solutions so obtained to find the resolution necessary to get accurate enough solutions.
In the latter test, on the other hand, we compare time-dependent solutions with different angular resolutions, employing time-dependent source terms.
We will explain how to set the source term for each test in the following.

We begin with the steady-state test, in which we are interested in the asymptotic states alone although we solve the time evolutions leading to them.
The time-independent source term that is deployed in this test and is denoted by $S_{\mathrm{steady}}$ is constructed as follows.
We take a neutrino distribution function at a certain spatial point at a certain postbounce time from a simulation result (see below for details) and interpolate it to grid points in the finer mesh.
The advection term for this distribution function, $A_i(f_B)$, is then evaluated not by the direct calculation, which is a bit cumbersome because of the spatial derivatives involved, but by using the following relation valid for steady states:
\begin{equation}
    A_i(f_B) (= - S_{\mathrm{steady}} ) = C_i[f_B].
    \label{eq::Source_term_steady_case}
\end{equation}

Strictly speaking, $f_B$ is not a steady solution but a snapshot of a time-dependent solution.
The source term obtained this way is an approximation to the actual (minus) advection term.
It does not matter for our purpose, however.
The reference distribution, $f_B$, can be anything as long as it is reasonably close to the actual distribution and gives an equally reasonable approximation to the advection term.
There is another merit with this treatment. The steady solution of Eq.~\eqref{eq::Boltz_with_source} for the source term given by Eq.~\eqref{eq::Source_term_steady_case} satisfies the following simple equation: 
\begin{equation}
    C_i[f_i] = C_i[f_B],
\label{eq::Collision_term_relation_steady_state}
\end{equation}
that is, $f_i$ should be equal to $f_B$ if the same resolution is used on both sides of this equation.
If the resolution is lower on the left hand side, $f_i$ will be deviated from $f_B$, the fact which we will exploit to estimate the errors induced by the low-angular resolution.
Note that the right hand side of this equation is given on the finest angular mesh for the left hand side in this test.

\begin{figure}
  \begin{minipage}{1\linewidth}
    \centering
    \includegraphics[width=1\textwidth]{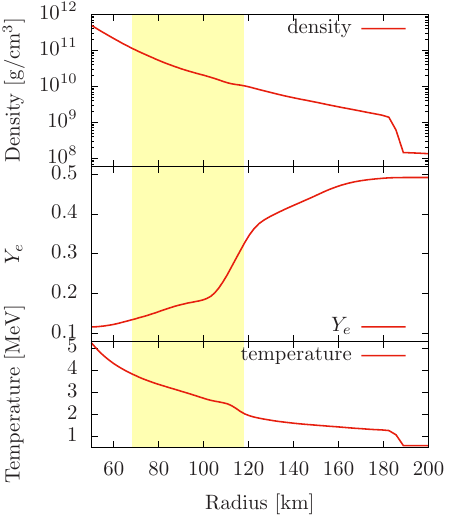}
    \subcaption{Radial profiles of background}
    
  \end{minipage}\\
  \caption{
  radial profiles of density $\rho$, electron fraction $Y_e$ and temperature $T$.
  Steady-state test and time-evolution test are performed at the edge and throughout the entire yellow-shaded region.
  }
  \label{fig:background_hydro}
\end{figure}

The reference neutrino distribution and matter properties are taken from our Boltzmann simulation of CCSN for a $15 M_\odot$ progenitor model \citep{Harada2022}.
This is a rapid rotation model.
In fact, we added by hand a nearly uniform rotation of a period of $4$ rad/s to the originally non-rotating stellar core.
This model is chosen because we want to investigate the resolution issue both for axisymmetric and non-axisymmetric neutrino distributions in momentum space.
We extract at $100$ms after bounce a matter profile along the radial ray at $\theta = \pi/4$, where the deviation from axisymmetry of the neutrino angular distribution in momentum space is expected to be largest.
It is shown in Fig.~\ref{fig:background_hydro}.
In this paper, we will focus on the yellow-shaded region, which spans the density range of $\rho = 10^{10} - 10^{11} \mathrm{g/cm^3}$. 
For the steady-state test, we perform the one-zone model calculations at the innermost point of the region ($\rho = 10^{11} \mathrm{g/cm^3} $).
For the time-evolution test, as we explain later, we construct the source term so that the neutrino transfer in this region could be mimicked.

The Boltzmann simulation, from which the reference neutrino distribution is constructed as above, has rather low-angular resolution $(10 (\theta_{\nu}) \times 6 (\phi_{\nu}))$.
We hence need first to interpolate the neutrino distribution obtained on the coarse grid to the grid points on much finer meshes so that the results could be used as the reference distributions indeed.
This is done as follows.
For the investigation of the $\theta_\nu$-resolution,  we first take an average of the original distribution function over $\phi_{\nu}$, making it axisymmetric, and then fit the resultant distribution with an $11$ th order Legendre function and the forward-peaked structure is represented.

For the study of the $\phi_\nu$-resolution and combined resolution, on the other hand, the original distribution taken from the Boltzmann simulation is approximated with the Legendre polynomials, $P_{l}$, up to the $11$th order in the $\theta_\nu$ direction for each $\phi_\nu$ grid point.
The resultant functions is evaluated at the $\theta_\nu$ grid points on the highest-resolution mesh.
The distribution functions in $\phi_\nu$ so obtained are then approximated with a Fourier series up to the $3$rd order and evaluated at the $\phi_\nu$ grid points on the highest-resolution mesh.
As mentioned earlier, the original model in the Boltzmann simulation is rotational and, as a consequence, the neutrino distribution extracted from it is non-axisymmetric but rather modestly.
In order to conduct the $\phi_\nu$-resolution test in more extreme conditions, we consider another model, in which we modify the distribution artificially as follows: utilizing the transformation of the solid angle under rotation, we incline the distribution function used for the $\theta_\nu$-resolution test toward the $z$-axis by $ \pi/4 $.
The numbers of angular grid points deployed in these test calculations, $(N_{\theta_\nu},N_{\phi_\nu})$, as well as those used for the reference distribution function, $(N_{\theta_\nu}^{\mathrm{src}},N_{\phi_\nu}^{\mathrm{src}})$, and the degrees of the polynomial for interpolation are listed in Table \ref{table::model_various_resolution}. 

\begin{table}
 \small
 \caption{ Models with various resolutions and reference models from steady-state and time-evolution tests }
 \centering
 \raggedright
 \noindent
 \label{table::model_various_resolution}
 \begin{minipage}{\linewidth}
 \raggedright
 \hspace{-10em}
 \scalebox{0.9}{
   \begin{tabular}{|c|c|c|c|c|c|c|c|}\hline
     \multicolumn{2}{|c|}{
       \begin{tabular}[c]{@{}c@{}}
         \hspace*{-18mm}
         \begin{picture}(90,12)
           \linethickness{0.2mm}
           \put(-3,12){\line(10,-1.2){123}}
           \put(78,4){resolution}
           \put(10,-1){model}
         \end{picture}
       \end{tabular}
      } & $N_{\theta_\nu}$ & $N_{\phi_\nu}$ & $N^{\mathrm{src}}_{\theta_\nu}$ & $N^{\mathrm{src}}_{\phi_\nu}$ & ${ N^{\mathrm{\scriptscriptstyle poly}}_{\theta_\nu} }$ \\
  \hhline{|=|=|=|=|=|=|=|}
  \multirow{18}{*}{ \hspace{-4em} 
  \makecell[c]{ \rotatebox{90}{Steady-state Test} } } & \multirow{4}{*}{$\theta_\nu$-resolution Test} & 10 & 6 & 100 & 6 & 9 \\
  & & 20 & 6 & 100 & 6 & 9 \\
  & & 40 & 6 & 100 & 6 & 9 \\
  & & 80 & 6 & 100 & 6 & 9 \\ \cline{2-7}
  & \multirow{4}{*}{ \hspace{-4em} \makecell[c]{$\phi_\nu$-resolution Test \\ (Boltzmann model) } } & 10 & 4 & 10 & 24 & $-$ \\ 
  & & 10 & 6 & 10 & 24 & $-$ \\
  & & 10 & 8 & 10 & 24 & $-$ \\
  & & 10 & 12 & 10 & 24 & $-$ \\ \cline{2-7}
  & \multirow{4}{*}{ \hspace{-4em} \makecell[c]{$\phi_\nu$-resolution Test \\ (inclined model) }} & 10 & 4 & 10 & 24 & $-$ \\ 
  & & 10 & 6 & 10 & 24 & $-$ \\
  & & 10 & 8 & 10 & 24 & $-$ \\
  & & 10 & 12 & 10 & 24 & $-$ \\ \cline{2-7}  
  & \multirow{3}{*}{ \hspace{-4em} \makecell[c]{Combined-resolution \\ Test \\ (Boltzmann model) } } & 5 & 3 & 40 & 24 & 6 \\ 
  & & 10 & 6 & 40 & 24 & 6 \\
  & & 20 & 12 & 40 & 24 & 6 \\  \cline{2-7}  
  & \multirow{3}{*}{ \hspace{-4em} \makecell[c]{Combined-resolution \\ Test \\ (inclined model) } } & 5 & 3 & 40 & 24 & 6 \\ 
  & & 10 & 6 & 40 & 24 & 6 \\
  & & 20 & 12 & 40 & 24 & 6 \\\hhline{|=|=|=|=|=|=|=|}
  \multirow{23}{*}{ \hspace{-4em} 
  \makecell[c]{ \rotatebox{90}{Time-evolution Test} } } & \multirow{5}{*}{ \makecell[c]{$\theta_\nu$-resolution \\ Test } } & 8 & 6 & 40 & 6 & 4 \\
  & & 10 & 6 & 40 & 6 & 2 \\
  & & 10 & 6 & 40 & 6 & 4 \\
  & & 10 & 6 & 40 & 6 & 9 \\
  & & 20 & 6 & 40 & 6 & 4 \\ \cline{2-7}
  & \multirow{4}{*}{ \hspace{-4em} \makecell[c]{$\phi_\nu$-resolution \\ Test \\ (Boltzmann model) } } & 10 & 4 & 10 & 24 & $-$ \\ 
  & & 10 & 6 & 10 & 24 & $-$ \\
  & & 10 & 8 & 10 & 24 & $-$ \\
  & & 10 & 12 & 10 & 24 & $-$ \\ \cline{2-7}
  & \multirow{4}{*}{ \hspace{-4em} \makecell[c]{$\phi_\nu$-resolution \\ Test \\ (inclined model) } } & 10 & 4 & 10 & 24 & $-$ \\ 
  & & 10 & 6 & 10 & 24 & $-$ \\
  & & 10 & 8 & 10 & 24 & $-$ \\
  & & 10 & 12 & 10 & 24 & $-$ \\ \cline{2-7}
  & \multirow{5}{*}{ \hspace{-4em} \makecell[c]{Combined-resolution \\ Test \\ (Boltzmann model) } } & 5 & 3 & 40 & 24 & 6 \\ 
  & & 10 & 6 & 40 & 24 & 6 \\
  & & 10 & 6 & 40 & 24 & 11 \\
  & & 20 & 12 & 40 & 24 & 6 \\
  & & 20 & 12 & 40 & 24 & 21 \\ \cline{2-7}
  & \multirow{5}{*}{ \hspace{-4em} \makecell[c]{Combined-resolution \\ Test \\ (inclined model) } } & 5 & 3 & 40 & 24 & 6 \\ 
  & & 10 & 6 & 40 & 24 & 6 \\
  & & 10 & 6 & 40 & 24 & 11 \\
  & & 20 & 12 & 40 & 24 & 6 \\
  & & 20 & 12 & 40 & 24 & 21 \\ \hhline{|=|=|=|=|=|=|=|}
  \end{tabular} }
\end{minipage}
\end{table}

We move on to the time-evolution test, in which we are interested in the validity of the dual-resolution strategy in the time-dependent situation.
For that purpose, the time-dependent source terms are constructed so that the resultant time evolution of the distribution function could follow the original change of the distribution along the radial stream line of neutrino in the region of our concern indicated by the yellow highlight in Fig.~\ref{fig:background_hydro}.
This may be understood as follows.
The (spatially finite-differenced) Boltzmann equation can be rewritten in terms of the Lagrange derivative $\frac{D}{Dt} = \frac{\partial}{\partial t} + c \frac{\partial}{\partial r}$ :
\begin{equation}
    \frac{1}{c} \frac{D f_i}{D t} + A^\prime_{i}[f_{i-1},f_{i},f_{i-1}] = C_i[f_i],
    \label{eq::Boltzmann_with_source_time_evolve}
\end{equation}
where the advection term is changed as $ A^\prime_{i}[f_{i-1},f_{i},f_{i-1}] = A_{i}[f_{i-1},f_{i},f_{i-1}] - c \left( \frac{\partial f}{\partial r} \right)_i$.
Then, as in the steady-state test, we move this modified advection term, $A^\prime_i$, to the right hand side of Eq.~\eqref{eq::Boltzmann_with_source_time_evolve} and replace it with the source term, $S^\prime_i$: 
\begin{equation}
    \frac{1}{c} \frac{D f_i}{D t} = C_i[f_i] + S^\prime_i .
    \label{eq::time_bolz_equation_with_source}
\end{equation}
In the equation, $S^\prime_i$ is known and local function of time.
As a result, all the equations thus obtained are decoupled from one another and can be solved independently as in the steady-state test.
The source terms are constructed again from the Boltzmann simulation data as explained below.

Equation~\eqref{eq::time_bolz_equation_with_source} is finite-differenced in time as
\begin{equation}
    \frac{1}{c} \frac{ f^{n+1} - f^{n} }{\Delta t} = C[f^{n+1} ] + {S}^n[f_B],
    \label{eq::time_bolz_equation_with_source_numerical}
\end{equation}
where the superscript $n$ denotes the time step and $\Delta t$ is the (constant) time increment between the consecutive time steps used in the test;
$f_B$ is the distribution function taken from the Boltzmann simulation (and interpolated to a finer mesh).
In this expression, we omit the spatial label $i$ intentionally, since we are considering in this test the time evolution that one would observe if one moved radially at light speed in the region in Fig. \ref{fig:background_hydro}.
The source term ${S}^n[f_B]$ is nothing but the (modified) advection term (with the sign reversed) that is evaluated $f_B$ at appropriate times and positions.
Using the Boltzmann equation that $f_B$ satisfies \footnote{Strictly speaking, the interpolated distribution does not satisfy the finite-differenced Boltzmann equation exactly.
This minor deviation is not a concern for the purpose of this paper, though.}, we rewrite it as
\begin{equation}
    {S}^n[f_B] = \frac{1}{c} \frac{D f_{B}}{Dt} - C[f_B] = \frac{1}{c}\frac{f_{B,j+1}^{n+1} - f_{B,j}^{n}}{\Delta t} - C[f_{B,j+1}^{n+1}],
    \label{eq::source_term_time_evolution}
\end{equation}
where the subscript $j$ specifies the grid point in the fictitious radial mesh, the cell width of which is equal to $c \Delta t$.
The time increment $\Delta t$ is the same as that in Eq.~\eqref{eq::time_bolz_equation_with_source_numerical} and is set to $\Delta t = 2 \times 10^{-7} \mathrm{s}$ in our tests.
Note that the second term on the right-most expression in Eq.~\eqref{eq::source_term_time_evolution}, i.e., the collision term for $f_B$, is evaluated at the $(n+1)$-th time step.
It is obvious that Eq.~\eqref{eq::source_term_time_evolution} is an extension of Eq.~\eqref{eq::Source_term_steady_case}.
By substituting Eq.~\eqref{eq::source_term_time_evolution} into Eq.~\eqref{eq::time_bolz_equation_with_source_numerical}, the equation can be cast into
\begin{equation}
    \frac{1}{c} \frac{ f^{n+1} - f^{n+1}_{B,j+1} }{\Delta t} = \frac{1}{c}\frac{f^{n} - f_{B,j}^{n}}{\Delta t} + C[f^{n+1} ] - C[f_{B,j+1}^{n+1}].
    \label{eq::effective_solve}    
\end{equation}
This indicates clearly that the difference between the calculated distribution, $f$, and the reference distribution, $f_{B}$, is originated from the difference in the angular resolutions used for the evaluation of the collision terms, i.e., the last two terms on the right hand side of Eq.~\eqref{eq::effective_solve}.

In this test, we solve Eq.~\eqref{eq::time_bolz_equation_with_source_numerical} with the source term given by Eq.~\eqref{eq::source_term_time_evolution}, varying the angular resolution in momentum space (see Table~\ref{table::model_various_resolution}) and compare the results to quantify the errors that the low-angular resolutions produce. 
The source term, Eq.~\eqref{eq::source_term_time_evolution}, is evaluated numerically from the same Boltzmann data.
Since the Boltzmann data are not provided uniformly in space, we interpolate them when need be.
Note that the angular resolution employed for the evaluation of $C[f_B]$ is the same as the highest resolution considered in this test.
It should also be obvious now why we evaluated $C[f_B]$ at the $(n+1)$-th time step.
Since the collision term, that is, the second from the last on the right hand side of Eq.~\eqref{eq::effective_solve} is stiff, it needs to be treated in the time-implicit manner.
As mentioned above, the source terms are designed to cancel the corresponding terms in Eq.~\eqref{eq::effective_solve}.

\section{Steady-state test}\label{sec:result}
In this section, we perform the one-zone calculations with the time-independent source term (see Sec. \ref{sec:validation}).
We first consider the $\theta_\nu$- and $\phi_\nu$-resolutions separately because they are qualitatively different.
We will investigate how many grid points are needed in each angular direction to evaluate the collision terms accurately for the typical neutrino distribution.
Then we explore the appropriate combination of angular resolutions by varying the number of grid points in the two angular directions simultaneously.
\subsection{$\theta_\nu$-Resolution Test} \label{steady_theta}
\begin{figure}
  \begin{minipage}{0.9\linewidth}
    \centering
    \includegraphics[width=1\textwidth]{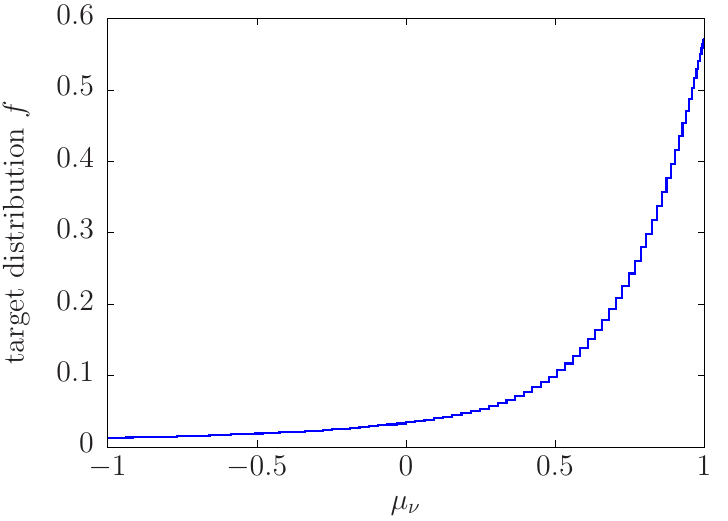}
    \subcaption{ $\epsilon = 1 \mathrm{MeV}$ }
  \end{minipage}\\
  \begin{minipage}{0.9\linewidth}
    \centering
    \includegraphics[width=1\textwidth]{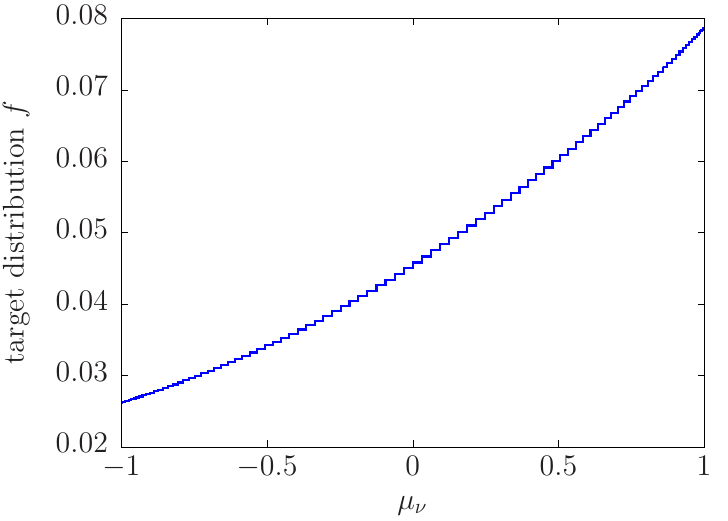}
    \subcaption{ $\epsilon = 11.1 \mathrm{MeV}$ }
  \end{minipage}\\
  \begin{minipage}{0.9\linewidth}
    \centering
    \includegraphics[width=1\textwidth]{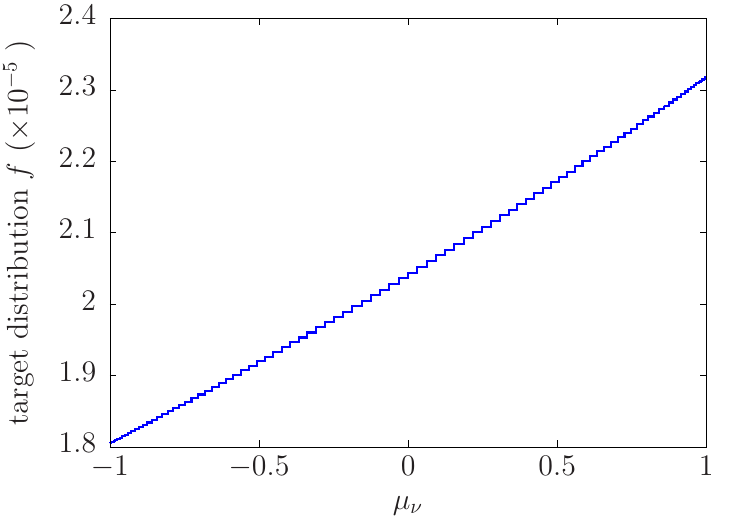}
    \subcaption{ $\epsilon = 41 \mathrm{MeV}$ }
  \end{minipage}
  \caption{ The target distributions with $( N_{\theta_\nu}^{\mathrm{scr}} , N_{\phi_\nu}^{\mathrm{scr}} ) = ( 100 , 6) $ for (a) $\epsilon = 1 \mathrm{MeV}$, (b) $\epsilon = 11.1 \mathrm{MeV}$ and (c) $\epsilon = 4 1 \mathrm{MeV}$.
  }
  \label{fig::steady_theta_target_distribution}
\end{figure}
\begin{figure}
  \begin{minipage}{0.8\linewidth}
    \centering
    \includegraphics[width=1\textwidth]{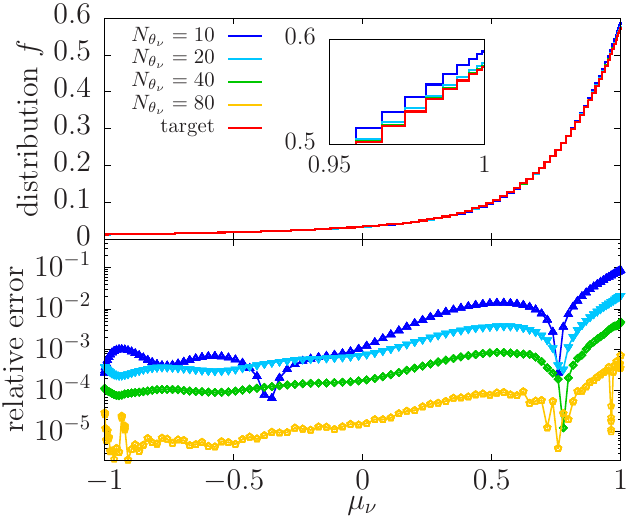}
    \subcaption{ $\epsilon = 1 \mathrm{MeV}$ }
  \end{minipage}\\
  \begin{minipage}{0.8\linewidth}
    \centering
    \includegraphics[width=1\textwidth]{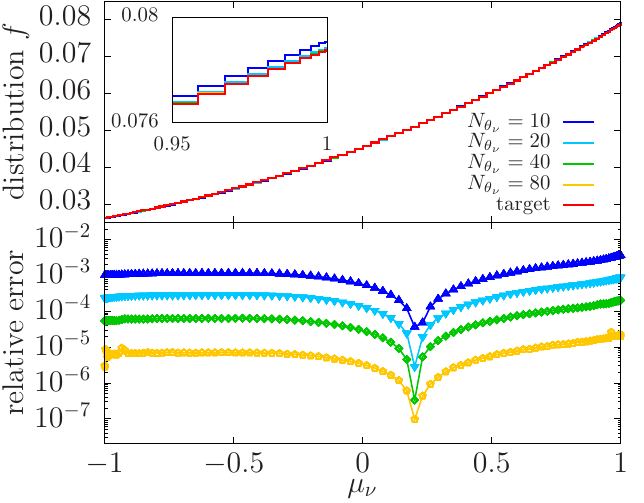}
    \subcaption{ $\epsilon = 11.1 \mathrm{MeV}$ }
  \end{minipage}\\
  \begin{minipage}{0.8\linewidth}
    \centering
    \includegraphics[width=1\textwidth]{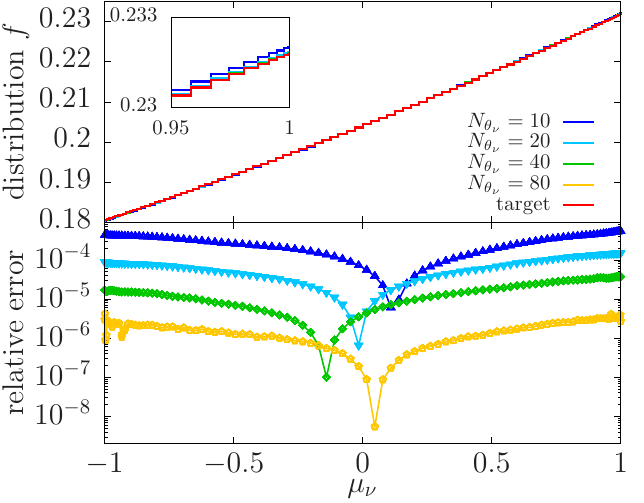}
    \subcaption{ $\epsilon = 41 \mathrm{MeV}$ }
  \end{minipage}
  \caption{
  The numerical solutions (upper half) and the relative errors (lower half) for (a) $\epsilon = 1 \mathrm{MeV}$, (b) $\epsilon = 11.1 \mathrm{MeV}$ and (c) $\epsilon = 4 1 \mathrm{MeV}$.
  The line color shows the angular resolution. 
  }
  \label{fig::steady_theta_target_and_numerical_solution}
\end{figure}
\begin{figure}
  \begin{minipage}{0.9\linewidth}
    \centering
    \includegraphics[width=1\textwidth]{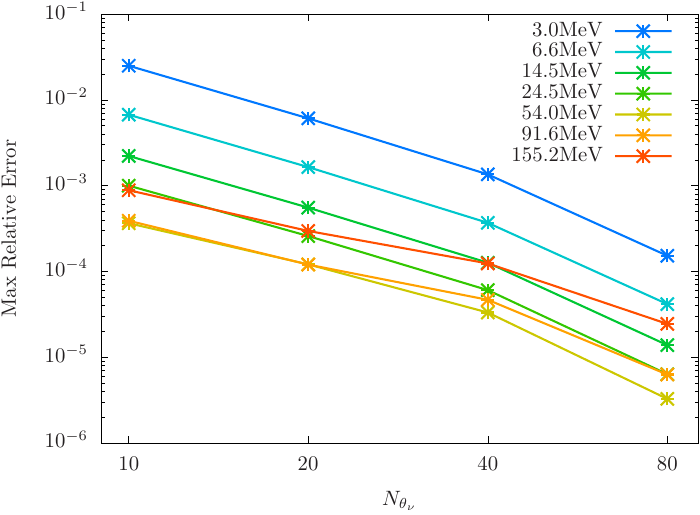}
    \subcaption{$\theta_\nu$-Resolution Test }
  \end{minipage}
  \caption{The relative errors in the steady-state $\theta_\nu$-resolution test.
  The reddish (bluish) lines correspond to lower (higher) energies.
  }
  \label{fig::theta_max_relative_error}
\end{figure}
\begin{figure*}
    \begin{minipage}{0.32\linewidth} 
        \centering
        \includegraphics[width=1\textwidth]{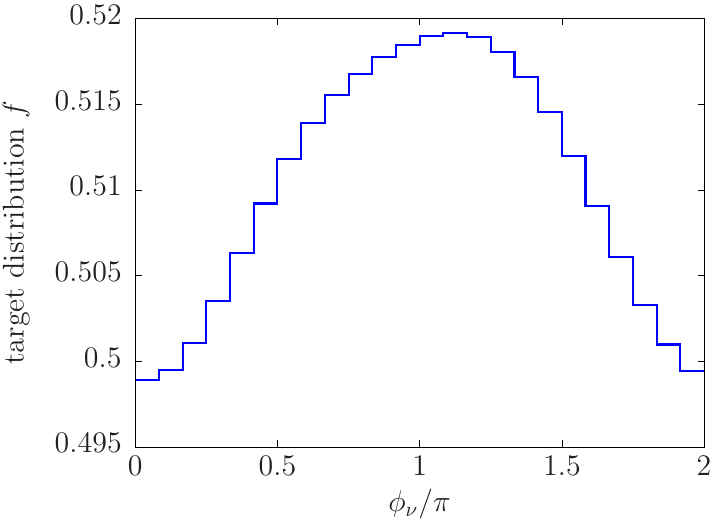}
        \subcaption{ Boltzmann model at $\epsilon = 1 \mathrm{MeV}$ } 
    \end{minipage}
    \hfill 
    \begin{minipage}{0.32\linewidth}
        \centering
        \includegraphics[width=1\textwidth]{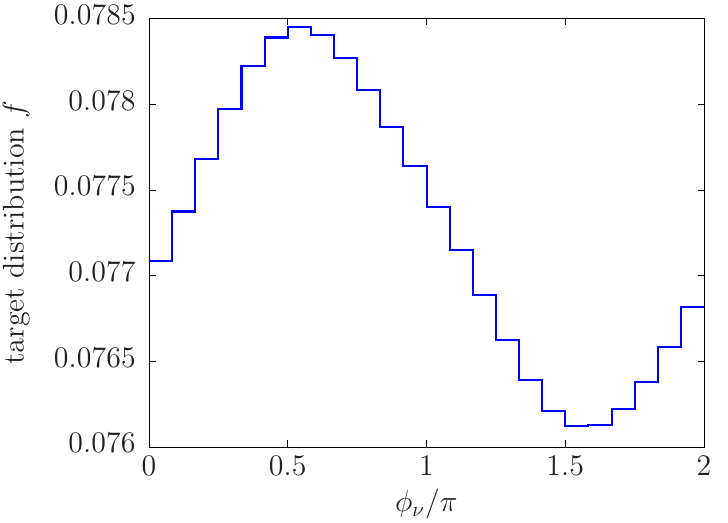}
        \subcaption{ Boltzmann model at $\epsilon = 11.1 \mathrm{MeV}$ }
    \end{minipage}
    \hfill
    \begin{minipage}{0.32\linewidth}
        \centering
        \includegraphics[width=1\textwidth]{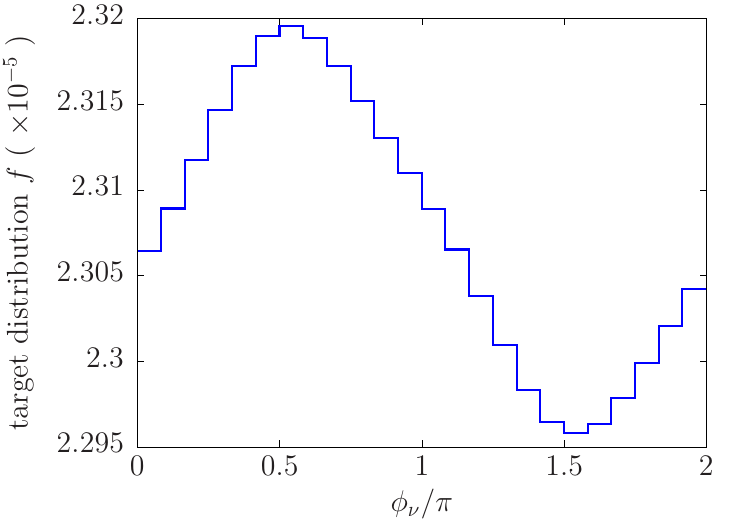}
        \subcaption{ Boltzmann model at $\epsilon = 41 \mathrm{MeV}$ }
    \end{minipage} \\ 
    \begin{minipage}{0.32\linewidth}
        \centering
        \includegraphics[width=1\textwidth]{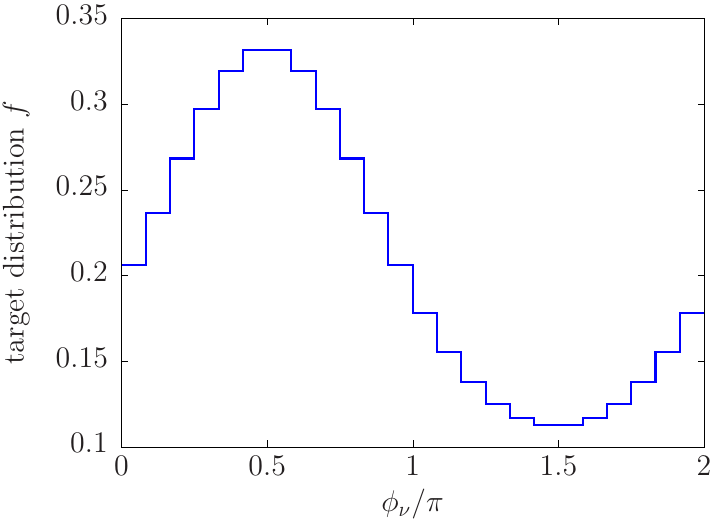}
        \subcaption{ Inclined model at $\epsilon = 1 \mathrm{MeV}$ }
        \label{fig::steady_inclined_1MeV_distribution}
    \end{minipage}
    \hfill
    \begin{minipage}{0.32\linewidth}
        \centering
        \includegraphics[width=1\textwidth]{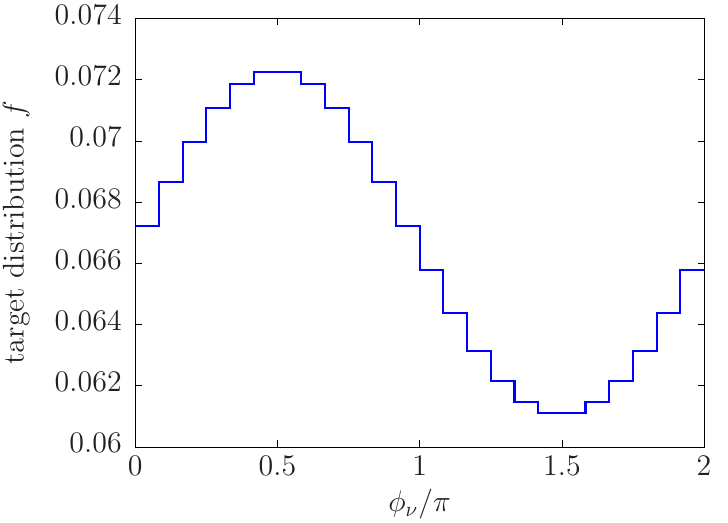}
        \subcaption{ Inclined model at $\epsilon = 11.1 \mathrm{MeV}$ }
        \label{fig::steady_inclined_11MeV_distribution}
    \end{minipage}
    \hfill
    \begin{minipage}{0.32\linewidth}
        \centering
        \includegraphics[width=1\textwidth]{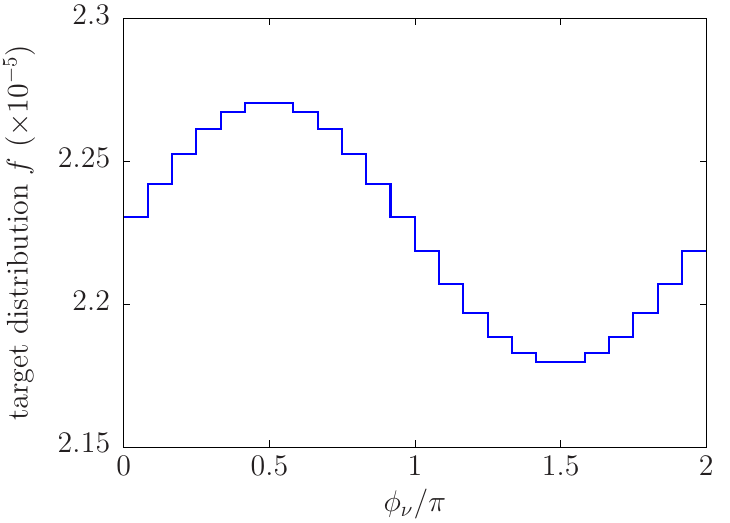}
        \subcaption{ Inclined model at $\epsilon = 41 \mathrm{MeV}$ }
        \label{fig::steady_inclined_41MeV_distribution}
    \end{minipage}
    \caption{
        The target distributions at $\cos\theta_\nu = 0.97$ (outward) as functions of $N_{\phi_\nu}$ for the Boltzmann model (upper panels) and the inclined model (lower panels) with $( N_{\theta_\nu}^{\mathrm{scr}} , N_{\phi_\nu}^{\mathrm{scr}} ) = ( 10 , 24)$.
        They are shown for three representative energies: $\epsilon = 1 \mathrm{MeV}$ (left column), $ 11.1 \mathrm{MeV}$ (middle column) and $4 1 \mathrm{MeV}$ (right column). 
    } \label{fig::target_distribution_steady_phi} 
\end{figure*}
\begin{figure*}
    \begin{minipage}{0.32\linewidth}
        \centering
        \includegraphics[width=1\textwidth]{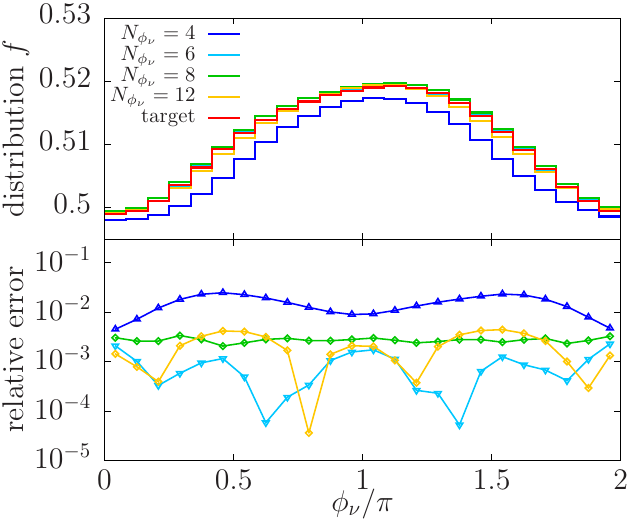}
        \subcaption{ $\epsilon = 1 \mathrm{MeV}$ }
        \label{fig::steady_Boltzmann_1MeV}
    \end{minipage}
    \hfill
    \begin{minipage}{0.32\linewidth}
        \centering
        \includegraphics[width=1\textwidth]{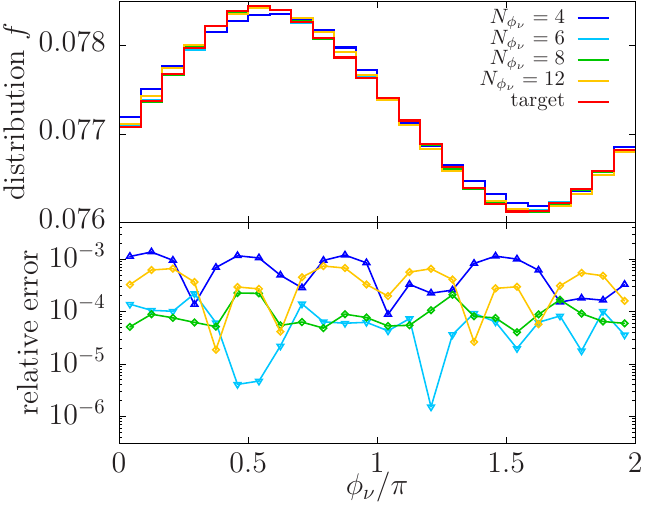}
        \subcaption{ $\epsilon = 11.1 \mathrm{MeV}$ }
        \label{fig::steady_Boltzmann_11MeV}
    \end{minipage}
    \hfill
    \begin{minipage}{0.32\linewidth}
        \centering
        \includegraphics[width=1\textwidth]{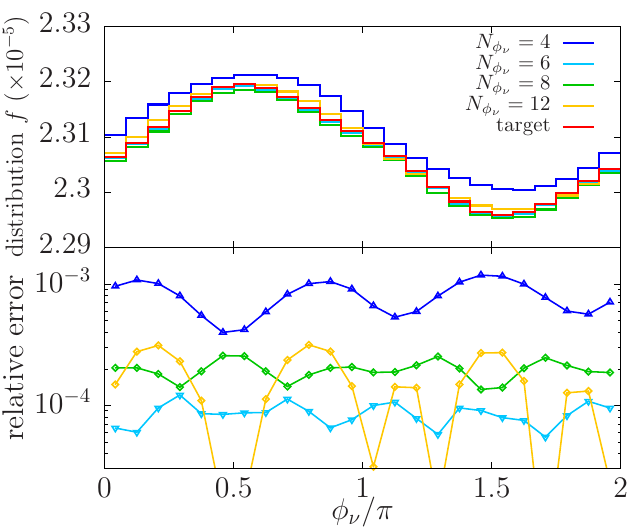}
        \subcaption{ $\epsilon = 41 \mathrm{MeV}$ }
        \label{fig::steady_Boltzmann_41MeV}
    \end{minipage}
    \begin{minipage}{0.32\linewidth}
        \centering
        \includegraphics[width=1\textwidth]{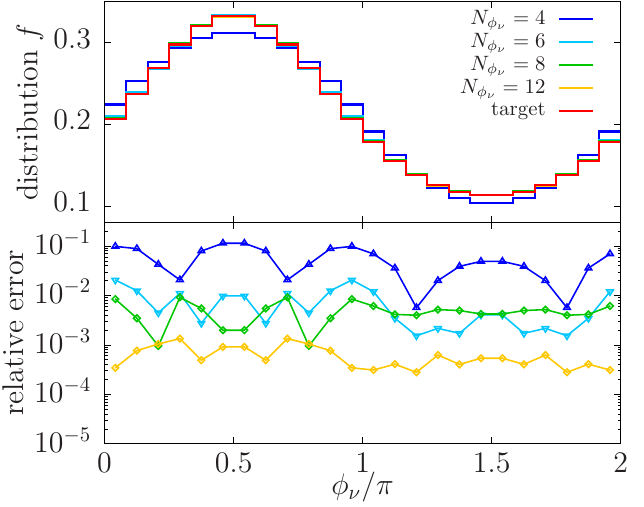}
        \subcaption{ $\epsilon = 1 \mathrm{MeV}$ }
        \label{fig::steady_inclined_1MeV}
    \end{minipage}
    \hfill
    \begin{minipage}{0.32\linewidth}
        \centering
        \includegraphics[width=1\textwidth]{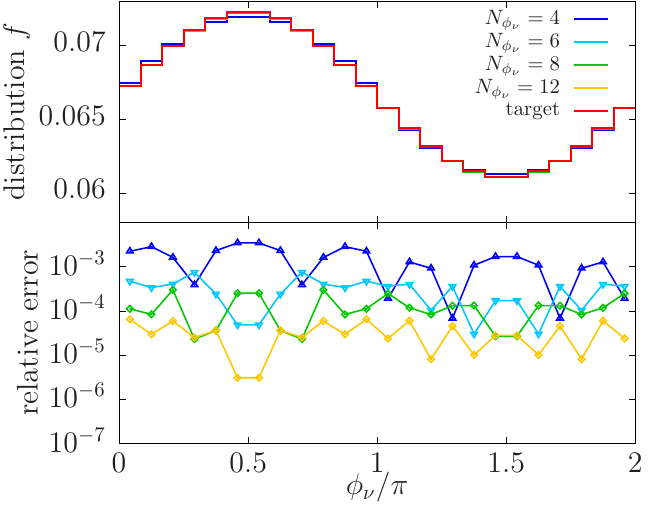}
        \subcaption{ $\epsilon = 11.1 \mathrm{MeV}$ }
        \label{fig::steady_inclined_11MeV}
    \end{minipage}
    \hfill
    \begin{minipage}{0.32\linewidth}
        \centering
        \includegraphics[width=1\textwidth]{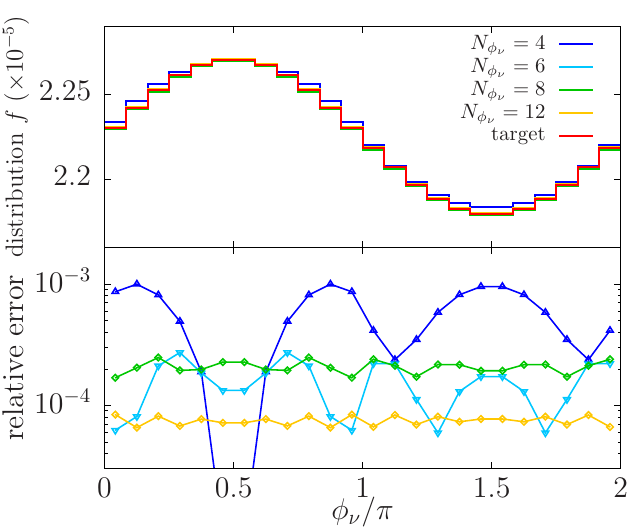}
        \subcaption{ $\epsilon = 41 \mathrm{MeV}$ }
        \label{fig::steady_inclined_41MeV}
    \end{minipage}
    \caption{
        The numerical solutions (upper half) and the relative errors (lower half) at $\cos\theta_\nu = 0.97$ for the Boltzmann model (upper panels) and the inclined model (lower panels) with $( N_{\theta_\nu}^{\mathrm{scr}} , N_{\phi_\nu}^{\mathrm{scr}} ) = ( 10 , 24)$.
        They are presented for three representative energies: $\epsilon = 1 \mathrm{MeV}$ (left column), $ 11.1 \mathrm{MeV}$ (middle column) and $ 4 1 \mathrm{MeV}$(right column). 
        The line color represents the angular resolution.
    }
    \label{fig::target_and_numerical_distribution_steady_phi} 
\end{figure*}
\begin{figure}
  \begin{minipage}{0.9\linewidth}
    \centering
    \includegraphics[width=1\textwidth]{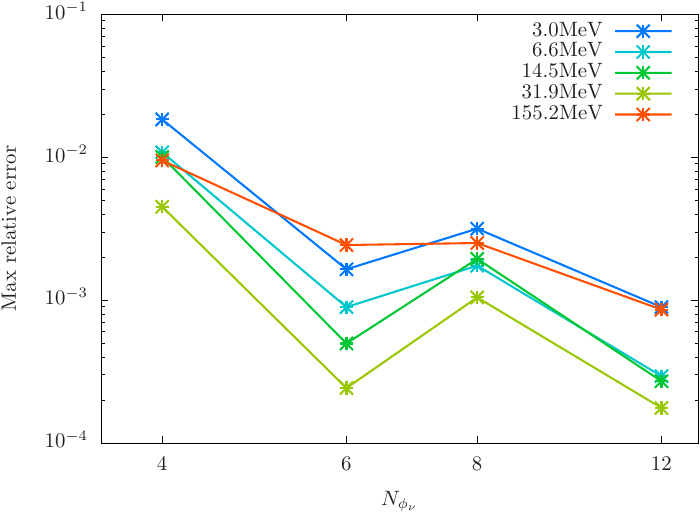}
    \subcaption{Boltzmann model}
     \label{fig::phi_rms_Boltzmann}
  \end{minipage}\\
  \begin{minipage}{0.9\linewidth}
    \centering
    \includegraphics[width=1\textwidth]{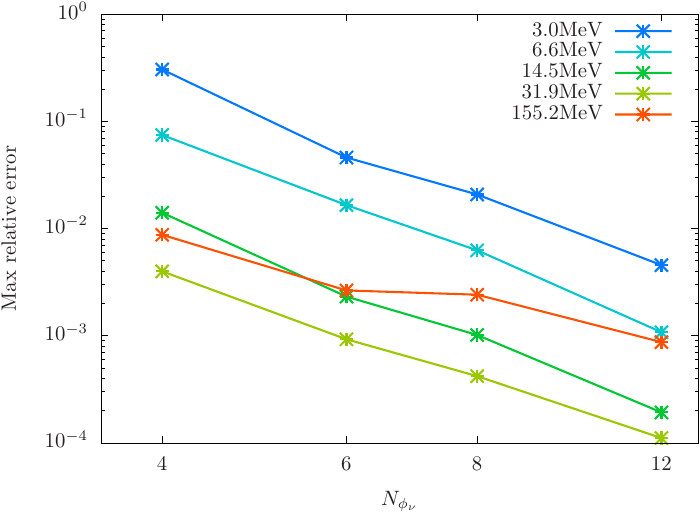}
    \subcaption{Inclined model}
     \label{fig::phi_rms_inclined}
  \end{minipage}
  \caption{The maximum relative errors as functions of $N_{\phi_\nu}$ in the steady-state $\phi_\nu$-resolution test for (a) the Boltzmann model and (b) the inclined model.
  The color coding is the same as in Fig. \ref{fig::theta_max_relative_error} .
  }
  \label{fig::steady_test_one_side}
\end{figure}
\begin{figure}
  \begin{minipage}{0.9\linewidth}
    \centering
    \includegraphics[width=1\textwidth]{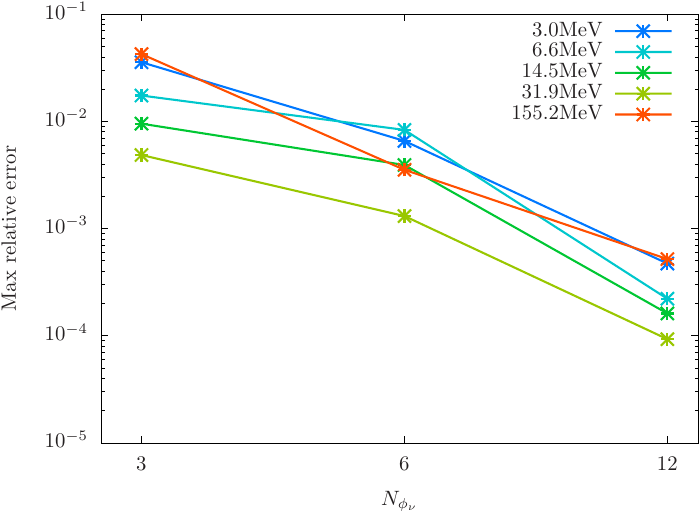}
    \subcaption{Boltzmann model in combined-resolution test}
     \label{fig::steady_angular_Boltzmann}
  \end{minipage}\\
  \begin{minipage}{0.9\linewidth}
    \centering
    \includegraphics[width=1\textwidth]{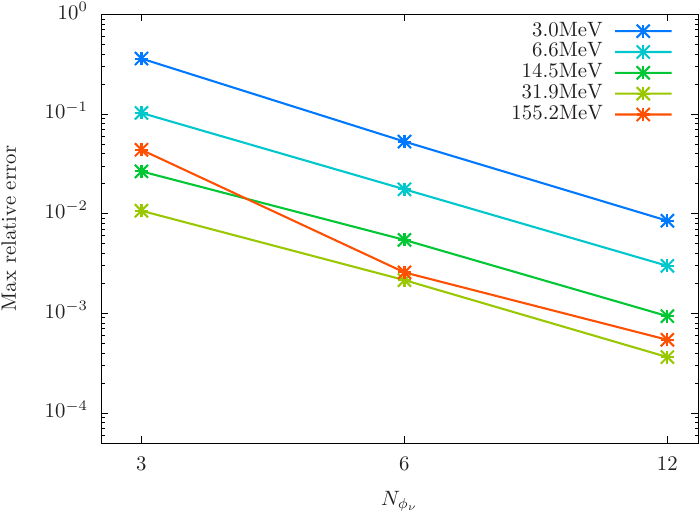}
    \subcaption{Inclined model in combined-resolution test}
     \label{fig::steady_angular_inclined}
  \end{minipage}
  \caption{Same as Fig.~\ref{fig::steady_test_one_side} but for the combined-resolution test.
  In the horizontal axis, $N_{\phi_\nu}=3$, $6$ and $12$ correspond to $(N_{\theta_\nu},N_{\phi_\nu}) = (5,3)$, $(10,6)$ and $(20,12)$, respectively.
  }
  \label{fig::steady_test_all_angle}
\end{figure}
In this test, we use an axisymmetric neutrino distribution in momentum space as a target, which is obtained by taking the average over $\phi_\nu$ of a snapshot of our Boltzmann simulation.
It is shown for three representative neutrino energies in Fig.~\ref{fig::steady_theta_target_distribution}. 
Note that this target distribution, $f_{\mathrm{target}}$, is represented with $100 (\theta_\nu) \times 6 (\phi_\nu)$ grid points.
We deploy the typical number, $6$, of $\phi_\nu$ grid points in this test although it has no influence on the result.
We perform the one-zone simulations, varying the number of grid points in $\theta_\nu$ from $10$ to $80$ (see Table \ref{table::model_various_resolution}).
By construction, we would get the target distribution exactly if we deployed $100$ grid points.

The numerical solutions are shown in Fig.~\ref{fig::steady_theta_target_and_numerical_solution}. 
Note that all of them use the $9$-th order polynomial, that is, $N^{\mathrm{poly}}_{\theta_\nu} = 9 $ in Eqs.~\eqref{eq::appr_distribution_theta_nu} and \eqref{eq:theta_poly}, to reconstruct the distribution function on the finer mesh from the one on the coarser mesh; this value of $N_{\theta_\nu}^{\mathrm{poly}}$ is also employed in building the target distribution.
We choose the same three different energies: one of them corresponds to the average energy whereas the other two are either higher or lower than this.
Since the scattering rate is proportional to the incident neutrino energy squared, the angular distribution is more forward-peaked for lower-energy neutrinos.
One finds that in all three cases, the results tend to converge to the target distribution as $N_{\theta_\nu}$ becomes larger.

In order to see this convergence quantitatively, we define the relative error, $\mathrm{err}(\epsilon,\theta_\nu,\phi_\nu,N_{\theta_\nu},N_{\phi_\nu})$, and its maximum, $\mathrm{err}_\mathrm{max} (\epsilon,N_{\theta_\nu},N_{\phi_\nu} )$ as
\begin{eqnarray}
    \mathrm{err}(\epsilon,  & & \theta_\nu, \phi_\nu,N_{\theta_\nu},N_{\phi_\nu}) \notag \\
    & & = \left|  \frac{f_{N_{\theta_\nu},N_{\phi_\nu}}(\epsilon,\theta_\nu,\phi_\nu) - f_\mathrm{target}(\epsilon,\theta_\nu,\phi_\nu)}{ \langle f_\mathrm{target}(\epsilon,\theta_\nu,\phi_\nu) \rangle } \right|, \label{eq::evaluate_error_in_steady_state}
\end{eqnarray}
\begin{equation}
    \mathrm{err}_\mathrm{max}(\epsilon,N_{\theta_\nu},N_{\phi_\nu}) = \max_{\theta_\nu,\phi_\nu}\{ \mathrm{err}(\epsilon,\theta_\nu,\phi_\nu,N_{\theta_\nu},N_{\phi_\nu}) \}, \label{eq::evaluate_max_error_in_steady_state}
\end{equation}
where $ \langle \cdot \rangle$ means the average over $\theta_\nu$ and  $\phi_\nu$.
We plot this quantity as a function of $N_{\theta_\nu}$ for several energies in Fig.~\ref{fig::theta_max_relative_error}.
As the angular resolution gets higher, the error becomes smaller monotonically for all energies. 
The error is largest at the lowest energy. 
This is because the neutrino distribution is the most anisotropic (see Fig.~\ref{fig::steady_theta_target_distribution}).
The relative error becomes larger again at very high neutrino energies, $\epsilon \gtrsim 90 \mathrm{MeV}$, simply because the population becomes smaller quickly at these energies.
We find that the maximum relative error is $\sim 2 \times 10^{-3}$ around the average neutrino energy ($\sim 15 \mathrm{MeV}$) even for $N_{\theta_\nu} = 10$, the resolution employed typically in our Boltzmann simulation. 
This suggests that $N_{\theta_\nu} = 10$ is sufficient for the collision term if the dual-resolution prescription is used.

\subsection{$\phi_\nu$-Resolution Test} \label{steady_phi}
For this test we adopt two target distribution functions of $\phi_\nu$ as described in Sec. \ref{sec:validation}.
One of them referred to as Boltzmann model in Table \ref{table::model_various_resolution} uses the distribution extracted from our Boltzmann simulation as it is.
Since this model is rotational, the angular distribution is non-axisymmetric.
The other named the inclined model is an extreme model, which exaggerates the $\phi_\nu$-dependence artificially.
They are displayed in Fig.~\ref{fig::target_distribution_steady_phi} as functions of $\phi_\nu$ for a fixed value of $\cos\theta_\nu = 0.97$.
We choose the same three neutrino energies as in Figs.~\ref{fig::steady_theta_target_distribution} and \ref{fig::steady_theta_target_and_numerical_solution}.
This time we deploy $24$ grid points in the $\phi_\nu$ direction for the target distribution.
We fix $N_{\theta_\nu}$ to $10$, the canonical number.

The distribution has one peak and one trough for both the Boltzmann and inclined models.
The degree of non-axisymmetry around the average energy is $ \sim 3 \%$ for the Boltzmann model but that for the inclined model is $\sim 20 \% $.
At the two other energies, non-axisymmetry of the inclined model is enhanced from that of the Boltzmann model by similar factors.
In most cases, the peak occurs at $\phi_\nu = \pi / 2$.
That is because neutrinos are dragged by rotating matters and the relativistic beaming occurs in the rotational direction.
Low energy neutrinos in the Boltzmann model (upper left panel in Fig.~\ref{fig::target_distribution_steady_phi}) are exceptional, having the peak at $\phi_\nu = \pi$.
This is because their neutrinospheres are smaller, making the relativistic beaming less important compared with the influence of the oblate density distribution, which tends to incline the neutrino distribution in the direction of the rotational axis; as a result, the peak occurs at $\phi_\nu = \pi$.

Varying $N_{\phi_\nu}$ from $4$ to $12$, we run the one-zone simulations.
The results are exhibited in Fig.~\ref{fig::target_and_numerical_distribution_steady_phi} for both models.
We choose the same three neutrino energies again.
For all energies and both models, numerical solutions become closer to the target distribution with $N_{\phi_\nu}$.
The relative error defined in Eq. \eqref{eq::evaluate_error_in_steady_state} with the target distribution replaced appropriately is evaluated for each run.
The results are plotted in Figs.~\ref{fig::phi_rms_Boltzmann} and \ref{fig::phi_rms_inclined}, respectively, for the Boltzmann model and the inclined model.
One finds again a general trend for both models that the error decreases with the number of the $\phi_\nu$ grid points.
The Boltzmann model with $N_{\phi_\nu} = 6$ is an exception, where the error is disproportionately small.
This is mainly because the original Boltzmann simulation was performed with this number of $\phi_\nu$ grid points.
It also contributes that one of the cell centers in this angular mesh with $N_{\phi_\nu}=6$ coincides with $\phi_\nu = \frac{\pi}{2}$, where the peak occurs in the distribution function.
We should emphasize that the errors at other resolutions in the Boltzmann (inclined) model obey a power law expected for the $\sim 3$rd ($\sim 4$th) order accuracy, which we believe better reflects the true performance of this method.

The errors tend to be larger both toward the highest and lowest neutrino energies in the Boltzmann model.
The reason is the same as in the $\theta_\nu$-resolution test: at very high energies, the neutrino population becomes smaller with the energy very rapidly whereas at very low energies, the neutrino distribution is highly anisotropic.
The inclined model has greater errors than the Boltzmann model in general.
This is just as expected, since the inclined model is designed so that the anisotropy should be artificially exaggerated.
Since the distribution function is almost isotropic at very high energies even in the inclined model, the errors are almost the same there between the two models.
The error around the average energy is $\lesssim 1 \%$ even in the inclined model for $ N_{\phi_\nu} = 6 $, the canonical number in our current Boltzmann simulation.
At lower energies, the error tends to be larger, $\lesssim 5 \%$ in the same extreme model. 
In the Boltzmann model, however, the error is smaller, $\lesssim 1 \%$, at all energies.
The canonical $\phi_\nu$-resolution will be hence reasonable for the collision term if it is combined with the dual-resolution prescription.

\subsection{Combined-Resolution Test} \label{steady_angular}
Finally we conduct the time-independent resolution test, varying both $N_{\theta_\nu}$ and $N_{\phi_\nu}$ simultaneously.
As in the $\phi_\nu$-resolution test, we use both the Boltzmann and inclined models.
In this test, the target distribution functions are built on the mesh with $(N_{\theta_\nu},N_{\phi_\nu})=(40,24)$.
We perform the one-zone calculations for $(N_{\theta_\nu},N_{\phi_\nu})=(5,3),(10,6)$ and $(20,12)$.
Note that $(N_{\theta_\nu},N_{\phi_\nu})=(10,6)$ is the resolution we normally use in our Boltzmann simulations of CCSNe.
For quantitative assessment, we evaluate the relative error defined in Eq.~\eqref{eq::evaluate_error_in_steady_state} with the target distribution replaced as in the $\phi_\nu$-resolution test.

The results are plotted in Figs.~\ref{fig::steady_angular_Boltzmann} and \ref{fig::steady_angular_inclined} for the Boltzmann model and the inclined model, respectively.
We take some representative energies as in Fig.~\ref{fig::steady_test_one_side}.
The results are as expected and the error obeys a power law of the $ \sim 1.4$-th ($\sim 1.25$-th) order in the Boltzmann (inclined) model.
Although we choose $N_{\phi_\nu}$ as the horizontal axis in these plots, the results are not changed if we use $N_{\theta_\nu}$ instead. 

The error is highest at the lowest neutrino energy, since the angular distribution is most anisotropic there. 
For the same reason the error is greater for the inclined model than for the Boltzmann model in general.



\section{Time-Evolution test} \label{sec:time_result}

\begin{figure}
  \begin{minipage}{0.9\linewidth}
    \centering
    \includegraphics[width=1\textwidth]{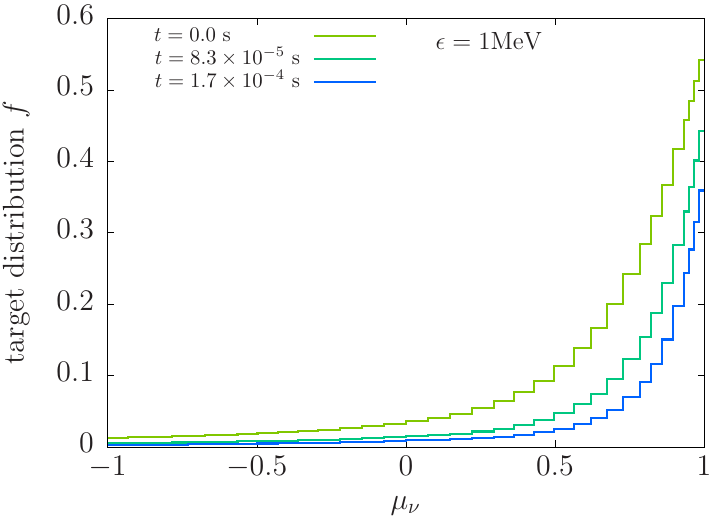}
    \subcaption{ $\epsilon = 1 \mathrm{MeV}$ }
  \end{minipage}\\
  \begin{minipage}{0.9\linewidth}
    \centering
    \includegraphics[width=1\textwidth]{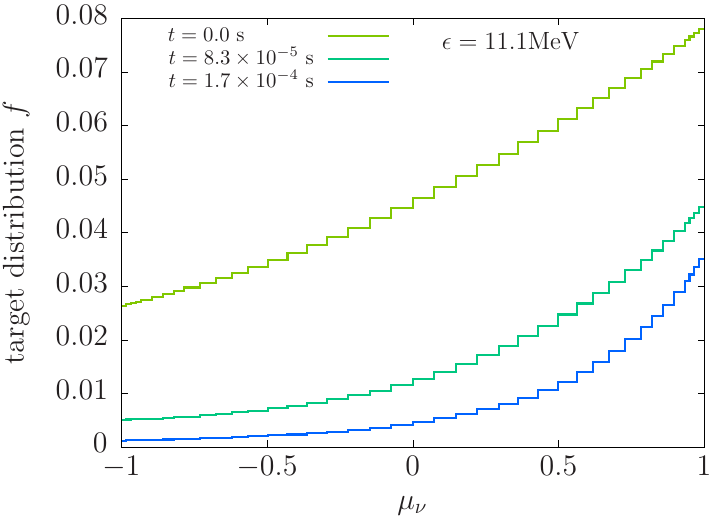}
    \subcaption{ $\epsilon = 11.1 \mathrm{MeV}$ }
  \end{minipage}\\
  \begin{minipage}{0.9\linewidth}
    \centering
    \includegraphics[width=1\textwidth]{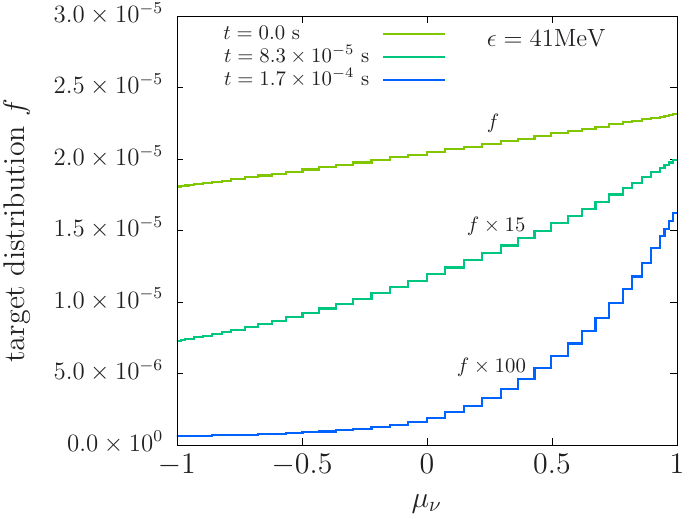}
    \subcaption{ $\epsilon = 41 \mathrm{MeV}$ }
  \end{minipage}
  \caption{ The target distributions for the time-evolution $\theta_\nu$-resolution test.
  They are presented at three different times for the neutrino energies of  (a) $\epsilon = 1 \mathrm{MeV}$, (b) $\epsilon = 11.1 \mathrm{MeV}$ and (c) $\epsilon = 4 1 \mathrm{MeV}$.
  We adopt $N_{\theta_{\nu}}^{\mathrm{src}} = 100$ and $N_{\phi_\nu}^{\mathrm{src}} = 6$.
  }
  \label{fig::theta_target_distribution_time_evolve}
\end{figure}
\begin{figure}
  \begin{minipage}{0.8\linewidth}
    \centering
    \includegraphics[width=1\textwidth]{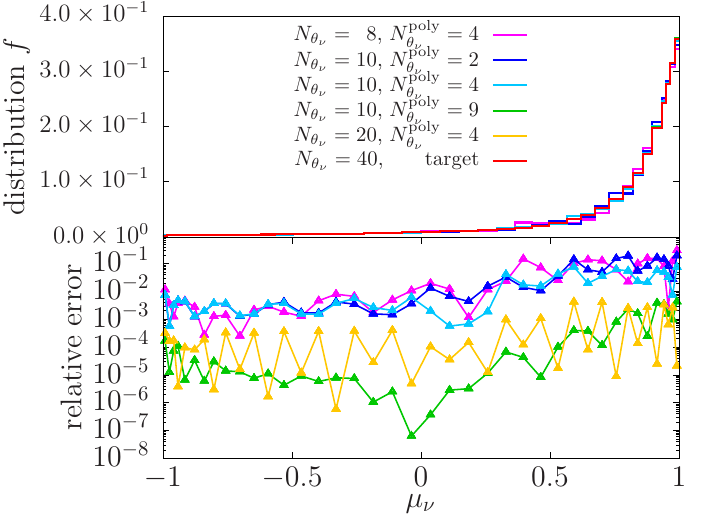}
    \subcaption{ $\epsilon = 1 \mathrm{MeV}$ }
    \label{fig::theta_target_and_numerical_solution_time_evolve_1MeV}
  \end{minipage}\\
  \begin{minipage}{0.8\linewidth}
    \centering
    \includegraphics[width=1\textwidth]{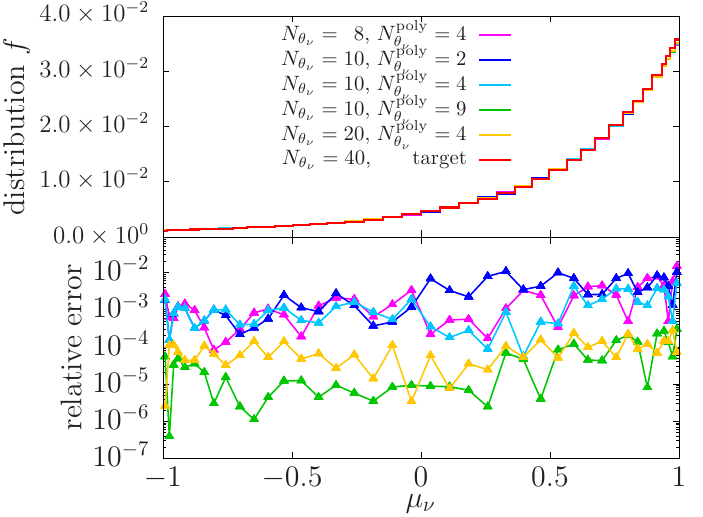}
    \subcaption{ $\epsilon = 11.1 \mathrm{MeV}$ }
    \label{fig::theta_target_and_numerical_solution_time_evolve_11MeV}
  \end{minipage}\\
  \begin{minipage}{0.8\linewidth}
    \centering
    \includegraphics[width=1\textwidth]{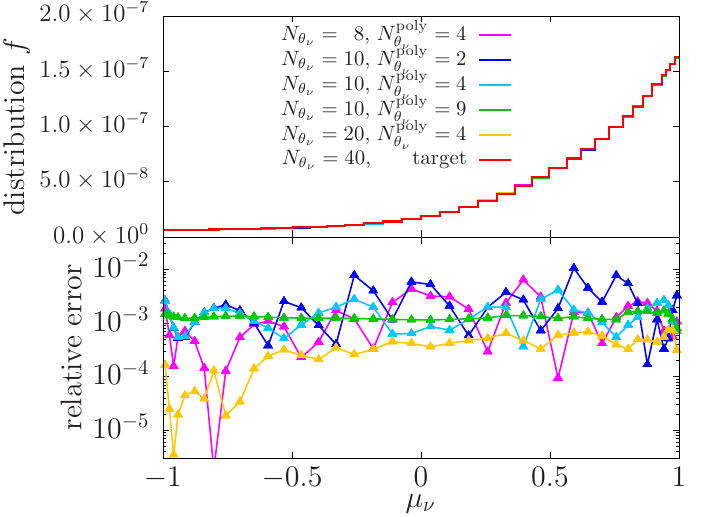}
    \subcaption{ $\epsilon = 41 \mathrm{MeV}$ }
    \label{fig::theta_target_and_numerical_solution_time_evolve_40MeV}
  \end{minipage}
  \caption{
  The numerical solutions (upper half) and the relative errors (lower half) as functions of $\mu_\nu$ in the time-evolution $\theta_\nu$-resolution test for the neutrino energies of (a) $\epsilon = 1 \mathrm{MeV}$, (b) $\epsilon = 11.1 \mathrm{MeV}$ and (c) $\epsilon = 4 1 \mathrm{MeV}$.
  }
  \label{fig::theta_target_and_numerical_solution_time_evolve}
\end{figure}

In this section, we perform one-zone calculations with the time-dependent source term (see Sec.\ref{sec:validation}).
We consider an evolution of the neutrino distribution function that would be seen by an imaginary observer moving radially outward with the speed of light in the yellow-highlighted region in Fig. \ref{fig:background_hydro}.
As in the steady-state test, the $\theta_\nu$- and $\phi_\nu$-directions are studied individually first; then the combined-resolution dependence is investigated.

\subsection{$\theta_\nu$-Resolution Test} \label{time_evo_theta}

\begin{figure}
  \begin{minipage}{0.8\linewidth}
    \centering
    \includegraphics[width=0.95\textwidth]{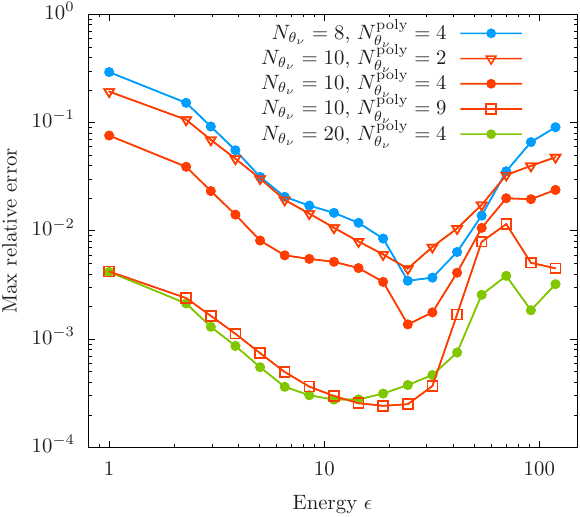}
    \subcaption{}
    \label{fig::relative_error_theta_interpolation}
  \end{minipage}\\
  \begin{minipage}{0.8\linewidth}
    \centering
    \includegraphics[width=0.95\textwidth]{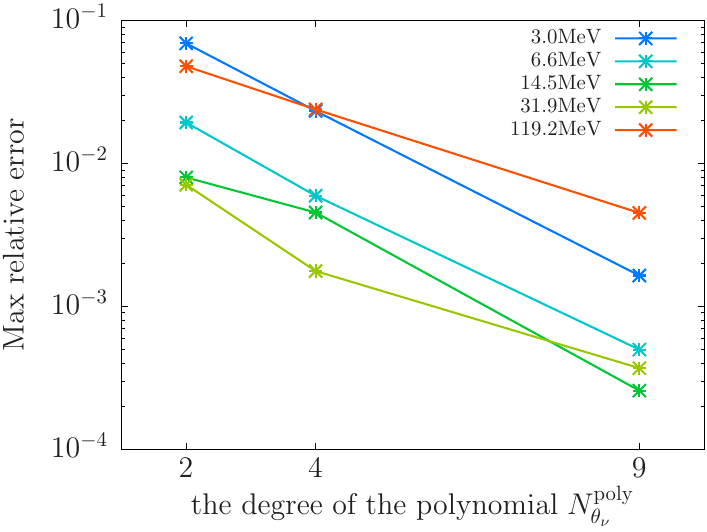}
    \subcaption{}
    \label{fig::relative_error_theta_test_steady_polynomial_dependence}
  \end{minipage}\\
  \begin{minipage}{0.8\linewidth}
    \centering
    \includegraphics[width=0.95\textwidth]{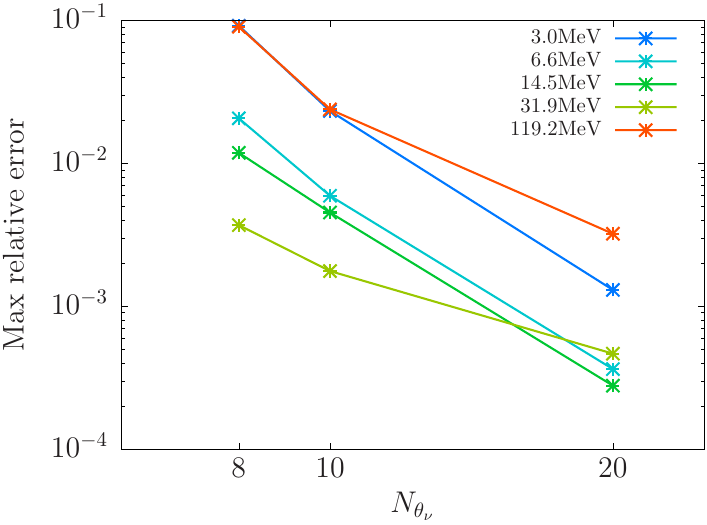}
    \subcaption{}
    \label{fig::relative_error_theta_test_steady_resoltuion_dependence}
  \end{minipage}
  \caption{
  The maximum relative errors as functions of (a) neutrino energy, (b) $N_{\theta_\nu}^{\mathrm{poly}}$ and (c) $N_{\theta_\nu}$ in the time-evolution $\theta_\nu$-resolution  test.
  $N_{\theta_\nu}$ is set to $10$ in panel (a) while $N_{\theta_\nu}^{\mathrm{poly}}$ is fixed to $4$ in panel (c).
  }
  \label{fig::relative_error_theta_test_in_steady_state}
\end{figure}

In this test, we use the axisymmetric neutrino distributions in momentum space, which are derived by taking the average over $\phi_\nu$ of the distributions at different radial positions of the same snapshot of our Boltzmann simulation as employed for the corresponding steady-state test.
The target distributions so obtained are represented with $40(\theta_\nu) \times 6(\phi_\nu)$ grid points and regarded as the time evolution.
We deploy the typical number, $6$, of $\phi_\nu$ grid points again although it has no influence on the result.
In Fig.~\ref{fig::theta_target_distribution_time_evolve}, we plot them for three representative energies at three different times. As the time passes (or as the imaginary observer moves outward), the neutrino distributions become more forward-peaked at all energies.

In this test, we investigate not only the dependence of results on $N_{\theta_\nu}$, the number of the grid points in the coarse mesh, but also the dependence on $N_{\theta_\nu}^{\mathrm{poly}}$, the degree of the polynomial used for interpolation.
We try $N_{\theta_\nu}^{\mathrm{poly}} = 2$ and $4$ in addition to the default choice of $9$.
The relative error is evaluated by Eq.~\eqref{eq::evaluate_error_in_steady_state} with the target distribution given at the final time of the calculation.
The results are plotted for the same three representative energies in Figs. \ref{fig::theta_target_and_numerical_solution_time_evolve} and \ref{fig::relative_error_theta_test_in_steady_state}. 
As $N_{\theta_\nu}$ increases, the relative errors decrease for all energy bins irrespective of $N_{\theta_\nu}^{\mathrm{poly}}$.
For example, the maximum relative errors around the average energy (see Fig.~\ref{fig::theta_target_and_numerical_solution_time_evolve_11MeV}) are $ \sim 2 \times 10^{-2}$, $ \sim 6 \times 10^{-3}$, $ \sim 3 \times 10^{-4}$ for $N_{\theta_\nu} = 8,10$ and $20$, respectively, at $N_{\theta_\nu}^{\mathrm{poly}}=4$.
As $N_{\theta_\nu}^{\mathrm{poly}}$ rises, on the other hand, the error becomes smaller again as demonstrated for $N_{\theta_\nu} = 10$.

The higher-order polynomials are particularly effective at low energies.
This is because the collision term becomes less important than the advection term and the distribution function is more forward-peaked as the neutrino energy gets lower.
At high energies, on the other hand, the neutrino distribution is almost isotropic in momentum space and can be represented well with low-order polynomials alone.
As a result, little improvement is observed in Fig.~\ref{fig::theta_target_and_numerical_solution_time_evolve_40MeV} even if higher-order polynomials are employed.
It should be also mentioned that rather high relative errors, $\sim 1 \%$, at $\epsilon \gtrsim 50 \mathrm{MeV}$ (see Fig.~\ref{fig::relative_error_theta_interpolation}) are mainly due to the small population of these high energy neutrinos.
The maximum errors obey power laws approximately for both $N_{\theta_\nu}$ and $N_{\theta_\nu}^{\mathrm{poly}}$ (see Figs.~\ref{fig::relative_error_theta_test_steady_polynomial_dependence} and \ref{fig::relative_error_theta_test_steady_resoltuion_dependence}): the power is roughly $\sim - 2$ and $\sim - 3.3$.

Based on these results, we judge that the combination of $N_{\theta_\nu} = 10$, the current canonical value in our Boltzmann simulation, with the dual-resolution prescription with $N_{\theta_\nu}^{\mathrm{poly}} = 9 $ will be a reasonable choice.

\subsection{$\phi_\nu$-Resolution Test} \label{time_evo_phi}
\begin{figure*}
    \begin{minipage}{0.32\linewidth} 
        \centering
        \includegraphics[width=1\textwidth]{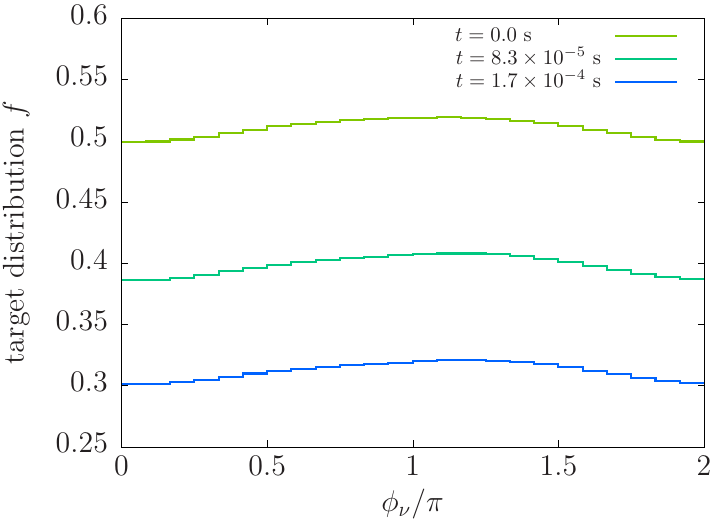}
        \subcaption{ Boltzmann model at $\epsilon = 1 \mathrm{MeV}$ } 
    \end{minipage}
    \hfill 
    \begin{minipage}{0.32\linewidth}
        \centering
        \includegraphics[width=1\textwidth]{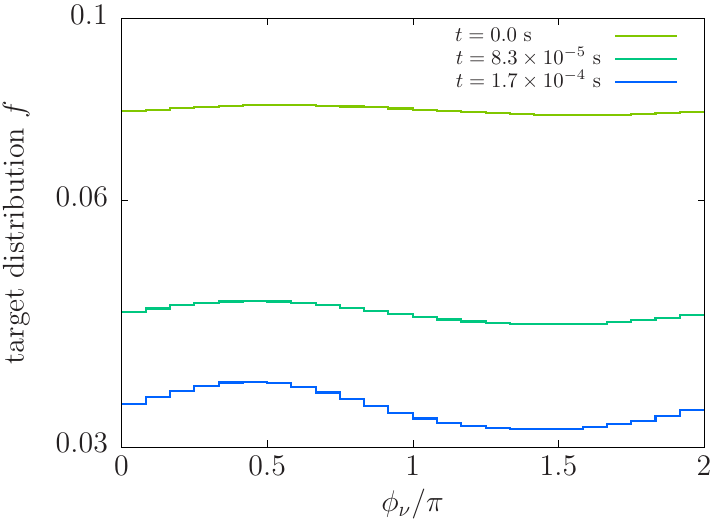}
        \subcaption{ Boltzmann model at $\epsilon = 11.1 \mathrm{MeV}$ }
    \end{minipage}
    \hfill
    \begin{minipage}{0.32\linewidth}
        \centering
        \includegraphics[width=1\textwidth]{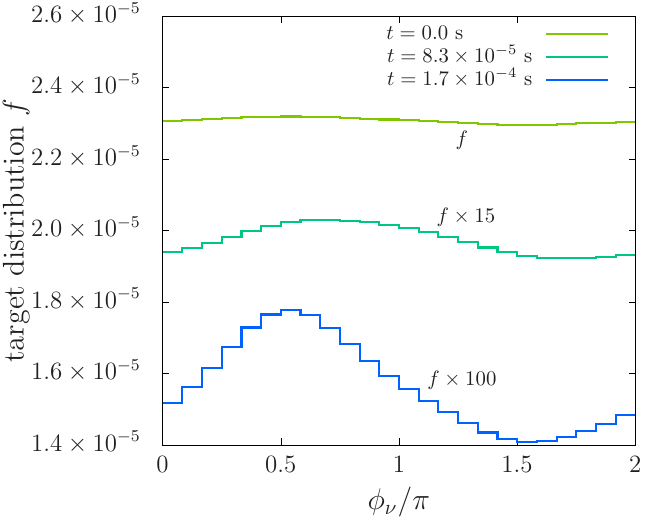}
        \subcaption{ Boltzmann model at $\epsilon = 41 \mathrm{MeV}$ }
    \end{minipage} \\ 
    \begin{minipage}{0.32\linewidth}
        \centering
        \includegraphics[width=1\textwidth]{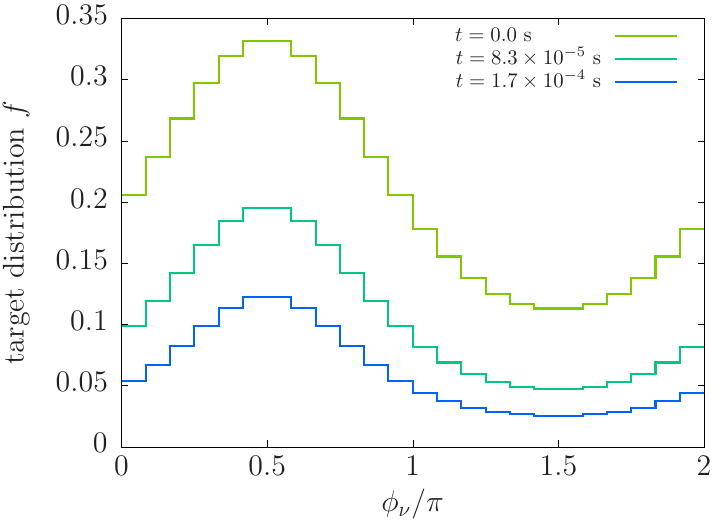}
        \subcaption{ Inclined model at $\epsilon = 1 \mathrm{MeV}$ }
        \label{fig::distribution_inclined_1MeV}
    \end{minipage}
    \hfill
    \begin{minipage}{0.32\linewidth}
        \centering
        \includegraphics[width=1\textwidth]{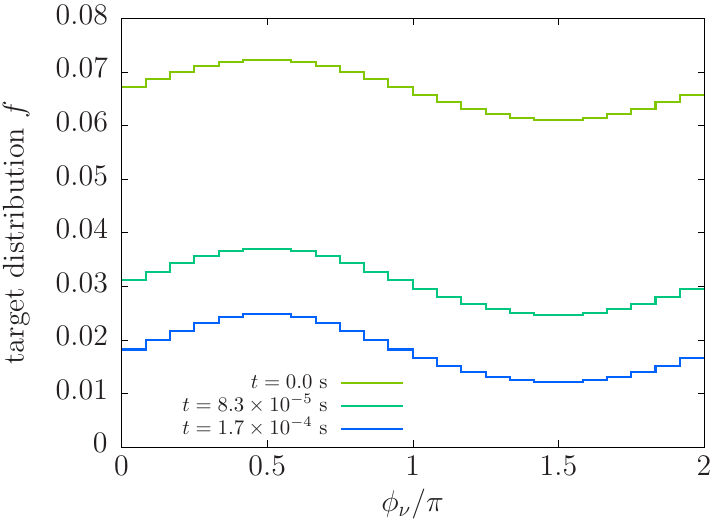}
        \subcaption{ Inclined model at $\epsilon = 11.1 \mathrm{MeV}$ }
        \label{fig::distribution_inclined_11MeV}
    \end{minipage}
    \hfill
    \begin{minipage}{0.32\linewidth}
        \centering
        \includegraphics[width=1\textwidth]{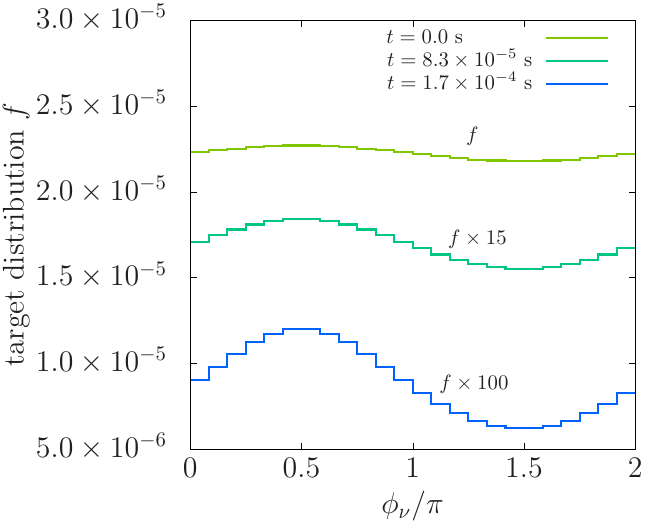}
        \subcaption{ Inclined model at $\epsilon = 41 \mathrm{MeV}$ }
        \label{fig::distribution_inclined_41MeV_time_evol}
    \end{minipage}
    \caption{
        The target distributions in $\phi_\nu$ at $\cos\theta_\nu = 0.97$ for the Boltzmann model (upper panels) and the inclined model (lower panels) with $( N_{\theta_\nu}^{\mathrm{scr}} , N_{\phi_\nu}^{\mathrm{scr}} ) = ( 10 , 24 )$.
        They are presented for three representative energies: $\epsilon = 1 \mathrm{MeV}$ (left column), $11.1 \mathrm{MeV}$ (middle column) and $ 41 \mathrm{MeV}$ (right column).
    } \label{fig::target_distribution_time_evol_phi} 
\end{figure*}
\begin{figure*}
    \begin{minipage}{0.32\linewidth}
        \centering
        \includegraphics[width=1\textwidth]{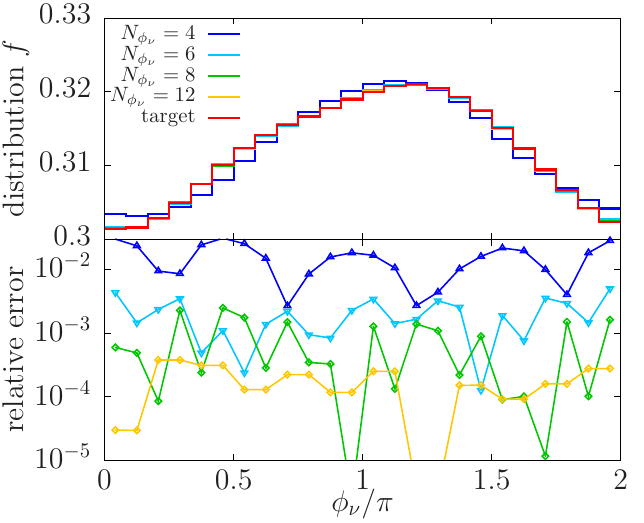}
        \subcaption{ $\epsilon = 1 \mathrm{MeV}$ }
        \label{fig::Boltzmann_1MeV_time_evolve}
    \end{minipage}
    \hfill
    \begin{minipage}{0.32\linewidth}
        \centering
        \includegraphics[width=1\textwidth]{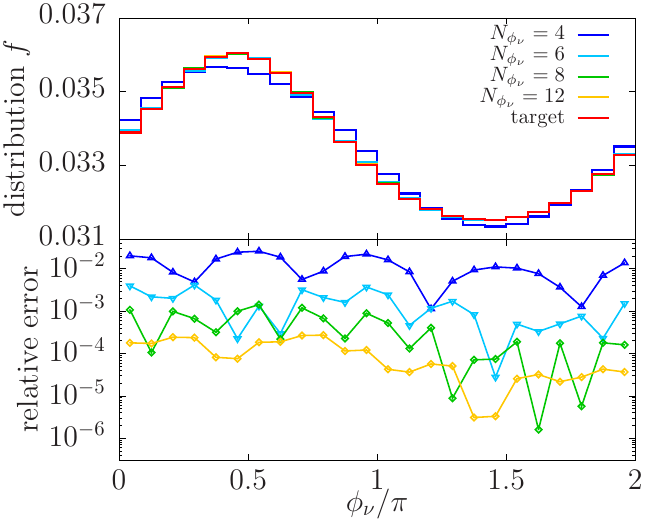}
        \subcaption{ $\epsilon = 11.1 \mathrm{MeV}$ }
        \label{fig::Boltzmann_11MeV_time_evolve}
    \end{minipage}
    \hfill
    \begin{minipage}{0.32\linewidth}
        \centering
        \includegraphics[width=1\textwidth]{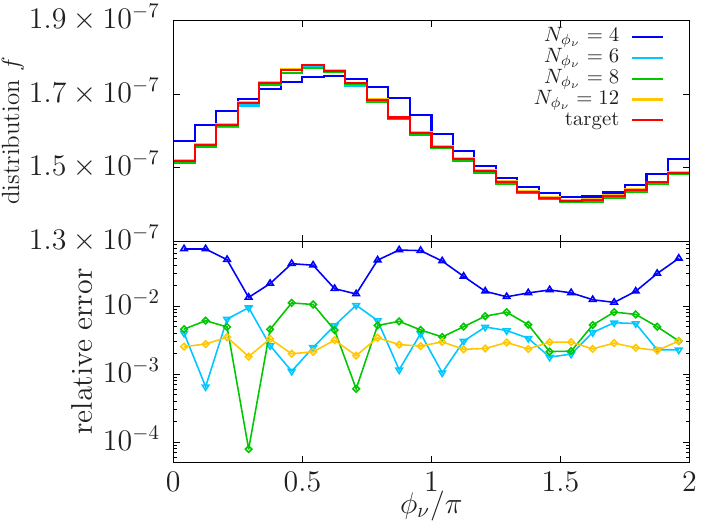}
        \subcaption{ $\epsilon = 41 \mathrm{MeV}$ }
        \label{fig::Boltzmann_41MeV_time_evolve}
    \end{minipage}
    \begin{minipage}{0.32\linewidth}
        \centering
        \includegraphics[width=1\textwidth]{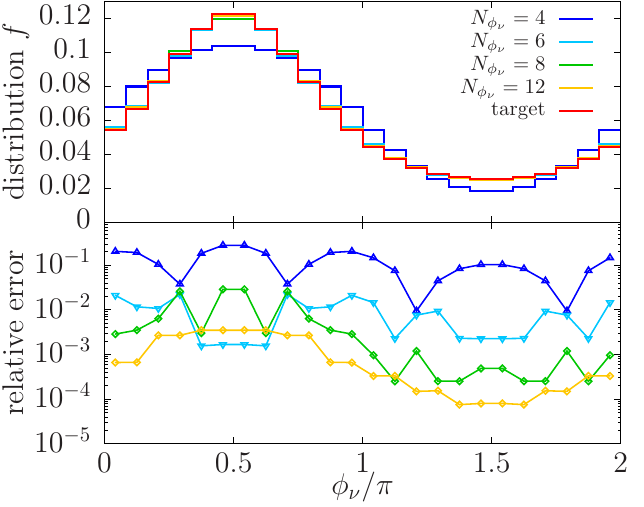}
        \subcaption{ $\epsilon = 1 \mathrm{MeV}$ }
        \label{fig::inclined_1MeV_time_evolve}
    \end{minipage}
    \hfill
    \begin{minipage}{0.32\linewidth}
        \centering
        \includegraphics[width=1\textwidth]{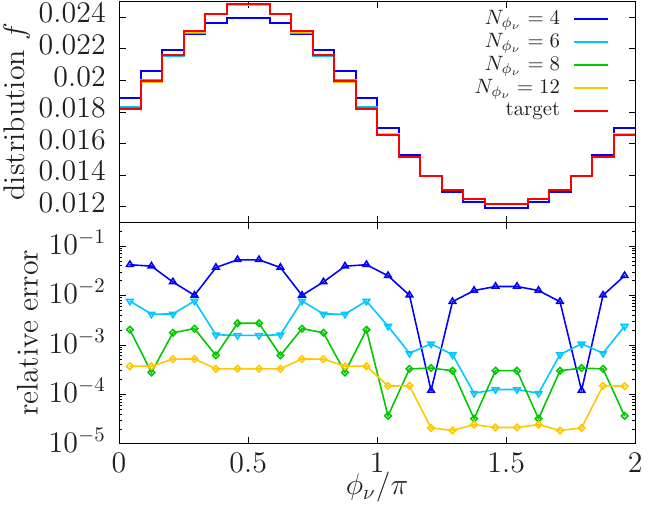}
        \subcaption{ $\epsilon = 11.1 \mathrm{MeV}$ }
        \label{fig::inclined_11MeV_time_evolve}
    \end{minipage}
    \hfill
    \begin{minipage}{0.32\linewidth}
        \centering
        \includegraphics[width=1\textwidth]{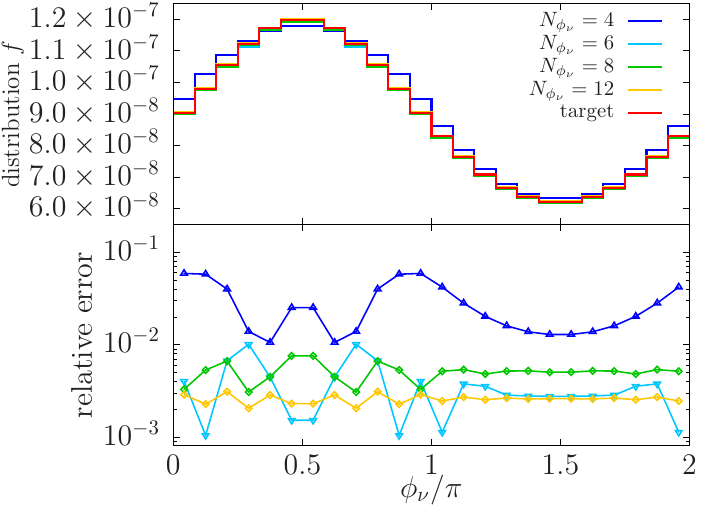}
        \subcaption{ $\epsilon = 41 \mathrm{MeV}$ }
        \label{fig::inclined_41MeV_time_evolve}
    \end{minipage}
    \caption{
        The numerical solutions (upper half) and the relative errors (lower half) at the end of simulations as functions of $\phi_\nu$ at $\cos\theta_\nu = 0.97$ for the Boltzmann model (upper panels) and the inclined model (lower panels) with $( N_{\theta_\nu}^{\mathrm{scr}} , N_{\phi_\nu}^{\mathrm{scr}} ) = ( 10 , 24)$.
        They are presented for the neutrino energies of $\epsilon = 1 \mathrm{MeV}$ (left column), $ 11.1 \mathrm{MeV}$ (middle column) and $4 1 \mathrm{MeV}$ (right column). 
        The line color represents the angular resolution.
    }\label{fig::target_and_numerical_distribution_time_evolve_phi} 
\end{figure*}

\begin{figure}
  \begin{minipage}{0.9\linewidth}
    \centering
    \includegraphics[width=1\textwidth]{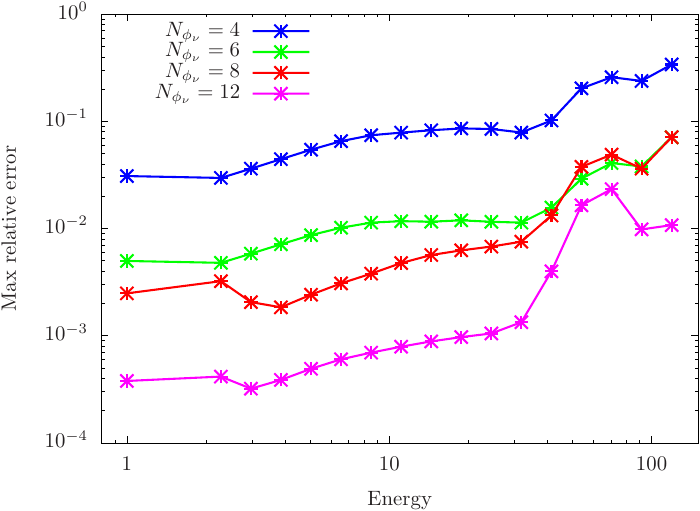}
    \subcaption{}
    \label{fig::relative_error_phi_Boltzmann}
  \end{minipage}\\
  \begin{minipage}{0.9\linewidth}
    \centering
    \includegraphics[width=1\textwidth]{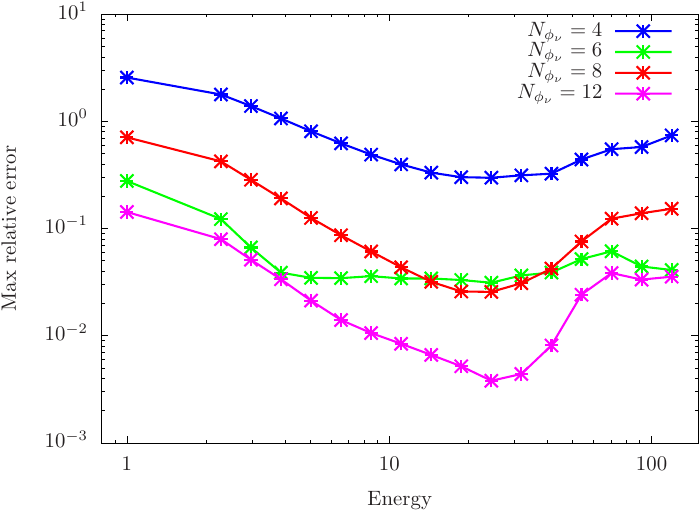}
    \subcaption{}
     \label{fig::relative_error_phi_Inclined}
  \end{minipage}
  \caption{
  The maximum relative errors as functions of neutrino energy in the time-evolution $\phi_\nu$-resolution test for (a) the Boltzmann model and (b) the inclined model.
  $N_{\theta_\nu}$ is fixed to $10$.
  The color denotes the value of $N_{\phi_\nu}$.
  The errors are evaluated at the final time step of each run.
  }
  \label{fig::relative_error_Nphi}
\end{figure}
\begin{figure}
  \begin{minipage}{0.9\linewidth}
    \centering
    \includegraphics[width=1\textwidth]{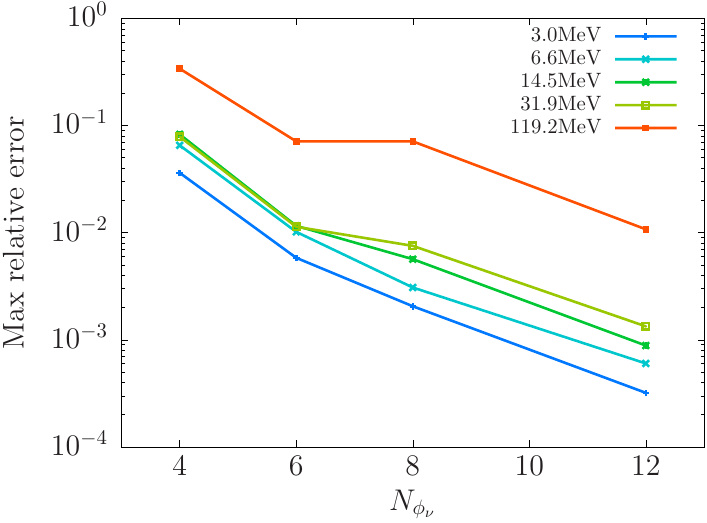}
    \subcaption{}
     \label{fig::phi_rms_Boltzmann_time_evol}
  \end{minipage}\\
  \begin{minipage}{0.9\linewidth}
    \centering
    \includegraphics[width=1\textwidth]{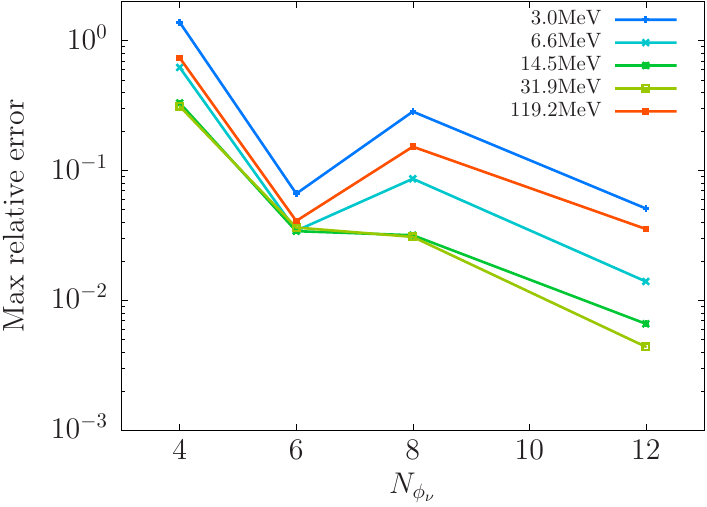}
    \subcaption{}
     \label{fig::phi_rms_inclined_time_evol}
  \end{minipage}
  \caption{The relative errors as functions of $N_{\phi_\nu}$ in the time-evolution $\phi_\nu$-resolution test for (a) the Boltzmann model and (b) the inclined model.
  The color shows the neutrino energy.
  }
  \label{fig::time_evol_test_one_side}
\end{figure}

For this test, we also adopt two target distribution functions of $\phi_\nu$: the Boltzmann model using one of our Boltzmann simulation data, and the inclined model with an artificially exaggerated anisotropy of $\phi_\nu$-dependence.
This time we introduce a time-dependence in the target distribution by moving an (imaginary) observer (see Sec.~\ref{sec:OneZone}).
They are displayed in Fig.~\ref{fig::target_distribution_time_evol_phi} for a fixed value of $\cos\theta_\nu=0.97$.
We choose the three representative values of neutrino energy $\epsilon$ as usual.
As the time passes, the distribution becomes more anisotropic for all energies.

In Fig.~\ref{fig::target_and_numerical_distribution_time_evolve_phi} we plot the numerical solutions and the relative errors as functions of $\phi_\nu$ at a fixed value of $\cos\theta_{\nu} = 0.97$ for the same three representative neutrino energies at the final time step.
As in the $\phi_\nu$-resolution test for the steady-state test, the numerical solutions tend to converge to the target distribution as $N_{\phi_\nu}$ increases, for all energies and in both models.
Around the average energy, the relative energy is less than $ 1\%$ for $N_{\phi_{\nu}} = 6$, the canonical number employed in our current Boltzmann simulations.
In the Boltzmann model, this is true for the other two energies.
In the inclined model, where the anisotropy is exaggerated artificially, the relative error is larger than in the Boltzmann model in general and at low and high energies in particular.
With $N_{\phi_{\nu}} = 8$, however, the relative error is still $\lesssim 1\%$ for the three energies.

The maximum relative errors are plotted for both models as functions of neutrino energy in Fig.~\ref{fig::relative_error_Nphi} and as functions of $N_{\phi_\nu}$ in Fig.~\ref{fig::time_evol_test_one_side}.
As the $\phi_\nu$ resolution increases, the relative errors decrease in general.
The errors for $N_{\phi_{\nu}}=6$ tend to be disproportionately small.
This is the same artifact that we observed earlier and that arises from the fact that this is the number of grid points employed in the original Boltzmann simulation, from which the target distributions are built.
The trend is more remarkable in the inclined model:
the relative errors for $N_{\phi_{\nu}} = 6$ are smaller than those for  $N_{\phi_{\nu}} = 8$.
This issue  is more apparent in the right panels, in which the relative errors are displayed as functions of $N_{\phi_{\nu}}$.

However, the general trend is clear.
At $N_{\phi_{\nu}} = 8$, the relative error is much less then $ 1\%$ around the average energy in the Boltzmann model;
it is a few $\%$ even for the inclined model, where the anisotropy is exaggerated artificially.
Again the relative error tends to be larger at very high energies both in the Boltzmann model and the inclined model.
The errors follow approximately the power law with the power of $\sim - 4$ .
One may notice that the energy dependence of the relative error for the lower energy region ($\lesssim20\mathrm{MeV}$) is opposite between the Boltzmann and the inclined model.
This is because the degree of anisotropy increases (decreases) with energy in the Boltzmann (inclined) model (see Fig. \ref{fig::target_and_numerical_distribution_time_evolve_phi}).
These results are consistent with the counterparts for the steady-state test.
Based on these results, we judge that $N_{\phi_{\nu}}=6$, the canonical value in our Boltzmann simulations, is not a bad choice but $N_{\phi_{\nu}}=8$ is a bit more preferred.

\subsection{Combined-Resolution test} \label{time_evo_angular}
\begin{figure}
  \begin{minipage}{0.9\linewidth}
    \centering
    \includegraphics[width=1\textwidth]{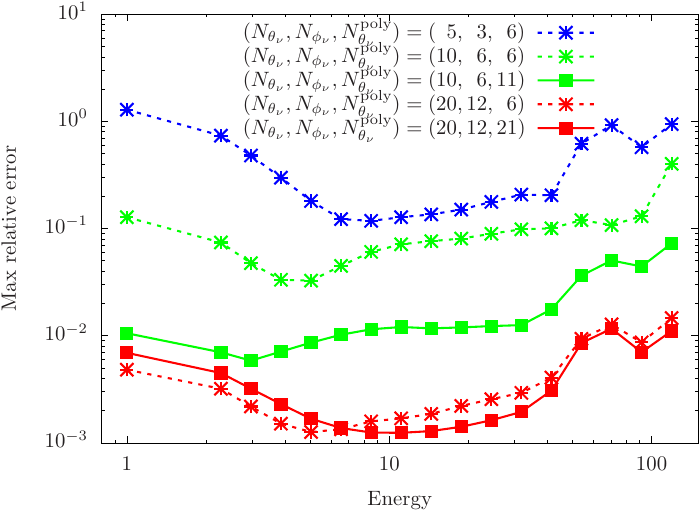}
    \subcaption{}
    \label{fig::relative_error_Boltzmann_all_angular}
  \end{minipage}\\
  \begin{minipage}{0.9\linewidth}
    \centering
    \includegraphics[width=1\textwidth]{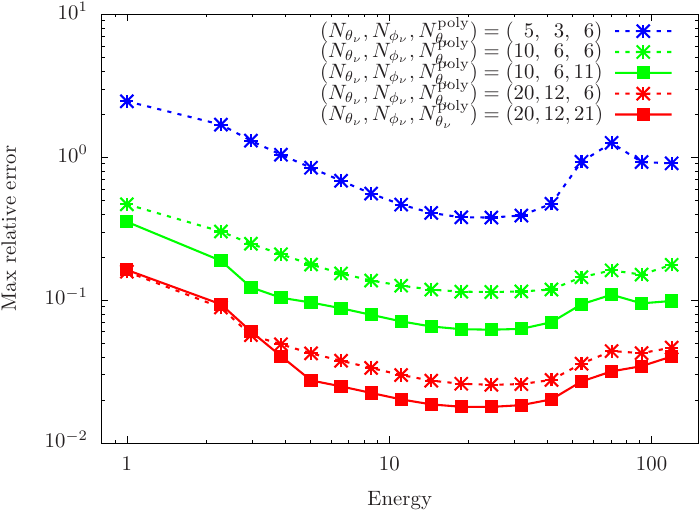}
    \subcaption{}
     \label{fig::relative_error_Inclined_all_angular}
  \end{minipage}
  \caption{
  The maximum relative errors as functions of neutrino energy in the time-evolution combined-resolution test for (a) the Boltzmann model and (b) the inclined model.
  The color shows different combinations of $N_{\theta_\nu}$, $N_{\phi_\nu}$ and $N_{\theta_\nu}^{\mathrm{poly}}$ as specified in the legend.
  }
  \label{fig::relative_error_all_angular}
\end{figure}
\begin{figure}
  \begin{minipage}{0.9\linewidth}
    \centering
    \includegraphics[width=1\textwidth]{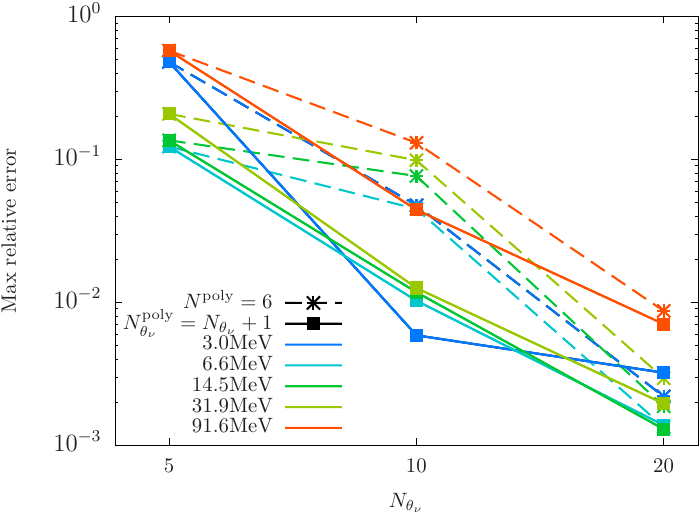}
    \subcaption{}
     \label{fig::time_evolve_combined_angular_Boltzmann}
  \end{minipage}\\
  \begin{minipage}{0.9\linewidth}
    \centering
    \includegraphics[width=1\textwidth]{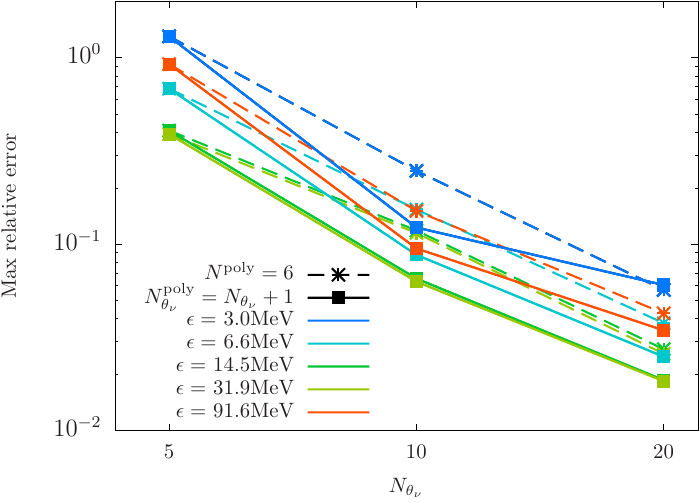}
    \subcaption{}
     \label{fig::time_evolve_combined_angular_inclined}
  \end{minipage}
  \caption{The maximum relative errors as functions of $N_{\phi_\nu}$ in the time-evolution combined-resolution test for (a) the Boltzmann model and (b) the inclined model.
  The color shows the neutrino energy whereas the line type distinguishes the polynomial order used in the $\theta_\nu$-interpolation:
  $N_{\theta_\nu}^{\mathrm{poly}}$ is fixed to $6$ for dashed lines whereas $N_{\theta_\nu}^{\mathrm{poly}} = N_{\theta_\nu}+1$ is adopted for solid lines.
  }
  \label{fig::time_evolve_combined_angular}
\end{figure}
As the last test of all, we study the accuracy of our dual-resolution prescription, varying both $N_{\theta_\nu}$ and $N_{\phi_\nu}$ (and $N_{\theta_\nu}^{\mathrm{poly}}$ as well) simultaneously for the same two target distributions as in  the $\phi_\nu$-resolution test.
The time dependence is also the same but they are constructed on the fine mesh with $(N_{\theta_\nu}^{\mathrm{src}},N_{\phi_\nu}^{\mathrm{src}}) = (40,24)$ this time.
The relative error is again measured by Eq.~\eqref{eq::evaluate_error_in_steady_state} for the distribution obtained at the final time of each run.
The results are plotted for both the Boltzmann model and the inclined model as functions of neutrino energy in Fig.~\ref{fig::relative_error_all_angular} and as functions of $N_{\phi_\nu}$ in Fig.~\ref{fig::time_evolve_combined_angular}.
The general trend is the same as in the previous two tests: the error decreases monotonically as the resolution becomes higher with $N_{\theta_\nu}^{\mathrm{poly}}$ fixed for both models; the error is larger for the inclined model, in which the anisotropy is amplified by hand.
The error is also reduced by increasing $N_{\theta_\nu}^{\mathrm{poly}}$ with $N_{\theta_\nu}$ and $N_{\phi_\nu}$ fixed although the reduction rate gets smaller as the angular resolution becomes higher.
Figure~\ref{fig::time_evolve_combined_angular} indicates again that the relative error in the Boltzmann (inclined) model obeys a power law with $\sim - 1.67$ ($\sim - 1.1$) as its power, which is a little bit different in modulus than the value observed in Fig.~\ref{fig::steady_test_all_angle} for the steady-state test.
We can see some peculiarities again at $N_{\phi_\nu} = 6$.

These results suggest that if the dual-resolution prescription is implemented, $(N_{\theta_\nu},N_{\phi_\nu})=(10,6)$, the current resolution employed in our Boltzmann simulations of CCSNe, will be acceptable although a bit larger number is preferable for $N_{\phi_\nu}$.
It seems that the higher the polynomial order is for the $\theta_\nu$ interpolation, the better the results are.
In fact, the best results are always obtained with $N_{\theta_\nu}^{\mathrm{poly}} = N_{\theta_\nu} + 1$, i.e., the entire interval $[-1,1]$ of $\mu_\nu = \cos\theta_\nu$ is expressed with a single high-order polynomial.

\section{Conclusion} \label{sec:Conclusion}
In order to address the issue of the angular resolution in momentum space in the $S_N$ method for Boltzmann neutrino transport, we propose a dual-resolution prescription and conduct some test calculations meant for proof of principle in this paper.
We treat the advection term and the collision term separately with different resolutions in the operator-splitting manner.
The distribution function in one of the resolutions is converted to the one in the other and vice versa.
In this paper, the first of a series of papers, we focus on the resolution appropriate for the collision term, since we already have some information on the resolution needed for the advection term \citep{richersDetailedComparisonMultiDimensional2017} and the treatment of the advection term in the high resolution is rather straightforward.
As the collision term, we consider only the neutrino scatterings on nucleons, one of the most important interactions in the supernova core, with its angular dependence and small energy transfer taken into account.
On the other hand, the advection term (and other interactions) is (are) taken from one of our Boltzmann simulations and treated as known quantities, which we call the source term.
Then the test calculations become local with different spatial points decoupled from one another.
We hence refer to these test calculations as the one-zone models in this paper.

The conversion of resolutions is non-trivial.
In fact, it is impossible to reproduce the distribution function in a higher resolution from the one given in a low resolution, mathematically speaking.
We hence seek for a way to reconstruct it with a reasonable accuracy.
We treat the zenith and azimuth directions separately and employ different polynomial interpolations for them.
This is because the functional behaviors are quite different between them.

We test the prescription with both time-independent and time-dependent source terms.
The former is modeled on a single point near the neutrinosphere whereas the latter is designed so that it should mimic the the advection term (and the interactions other than the neutrino-nucleon scattering combined) viewed by the imaginary observer moving radially outward at the light speed.

In both tests, we find a convergence of the numerical solutions to the exact solution as the number of grid points increases.
For the zenith-angle, $\theta_{\nu}$, direction, the higher order polynomials give better results whereas for the azimuth-angle, $\phi_{\nu}$, direction the quadratic functions deployed locally and combined together smoothly serve our purpose well.
In both directions the error follows some power-laws.
We judge that the angular mesh with $10 (\theta_{\nu}) \times 6 (\phi_{\nu})$, which is the grid used normally in our Boltzmann simulations, is reasonable for the calculation of the collision term if the dual-resolution prescription proposed in this paper is employed.

This study is just the first step mainly meant for the proof of principle of the dual-resolution prescription.
Now that we find the angular resolution suitable for the collision term and confirm the resolution conversion works well, in the next step we will consider the advection term squarely to confirm the conclusion of this paper in real neutrino transfer simulations.
In so doing, we will also implement all other interactions in the collision term; more elaborate schemes will be tried for the operator splitting of the advection and collision terms.
We find that errors tend to become larger at low energies.
This is mainly due to repeated resolution-conversions despite the collision term contributes very little and the same situation is expected for higher energies at larger radii.
In such situations, the collision term may be calculated less frequently or only the increment by the collision term should be interpolated to the finer mesh.
These possibilities will be also tested next.

\section{acknowledgments}
We thank Yudai Suwa for fruitful discussions.
This work used high-performance computing resources provided by Fugaku supercomputer at RIKEN, the Wisteria provided by JCAHPC through the HPCI System Research Project (Projects ID No. 240041, No. 240079, No. 240219, No. 240264, No. 250006, No. 250166, No. 250191, and No. 250226, JPMXP1020200109, JPMXP1020230406), Yukawa Institute of Theoretical Physics, the FX1000 provided by Nagoya University, Cray XC50 and XD2000 at the National Astronomical Observatory of Japan (NAOJ), the Computing Research Center at the High Energy Accelerator Research Organization (KEK), and Japan Lattice Data Grid (JLDG) on Science Information Network (SINET) of National Institute of Informatics (NII).
R.Y. is supported by JSPS KAKENHI Grant No. JP24K00632. H.N. is supported by Grant-in-Aid for Scientific Research (23K03468). S.Y. is supported by Grant-in-Aid for Scientific Research (25K01006).

\bibliography{sample631}
\bibliographystyle{aasjournal}

\appendix
\twocolumngrid
\section{Collision Terms and Energy-Subgrid}\label{sec:subgrid}
In this appendix we describe the treatment of collision terms.
In our dual-resolution formulation, they are handled with the low-resolution angular mesh.
Although there are various channels, through which neutrinos interact with matter, we focus on the scattering on nucleon, the dominant reaction that has non-trivial angular dependence.
In particular, we explain how to deal with small energy transfer between neutrino and nucleon in this almost iso-energetic scattering.

The collision term for the scattering process in the Boltzmann equation is expressed in general as
\begin{eqnarray} \label{eq:BoltzmannEquation}
C[f] & = & \displaystyle \int \frac{d{\epsilon^\prime}{\epsilon^\prime}^2}{(2\pi)^2}  d{\Omega^\prime} \left(R_{\mathrm{scat}}(\epsilon^\prime,\Omega^\prime;\epsilon,\Omega)  f(\epsilon^\prime,\Omega^\prime)[1 - f(\epsilon,\Omega)]\right.\nonumber\\
\qquad &-& \left. R_{\mathrm{scat}} (\epsilon,\Omega;\epsilon^\prime,\Omega^\prime)
f(\epsilon,\Omega)[1 - f(\epsilon^\prime,\Omega^\prime)]\right),
\end{eqnarray}
where $f$ is the neutrino distribution, $R_{\mathrm{scat}}(\epsilon,\Omega;\epsilon^\prime,\Omega^\prime)$ is the reaction kernel of the scattering $(\epsilon,\Omega) \xrightarrow{}
(\epsilon^\prime,\Omega^\prime)$, where $\epsilon$ and  $\epsilon^\prime$ are the energies of the incident and outgoing neutrinos, respectively, and $\Omega$ and $\Omega^\prime$ are the corresponding solid angles in momentum space.

The reaction kernel is taken from \cite{sugiuraLeptonicSemileptonicNeutrino2022}, neglecting the weak magnetism and other momentum transfer-dependent form factors in the weak currents of nucleon for simplicity.
Reinstating the omitted effects poses no difficulty.
Note also that the reaction rate adopted in this study incorporates the effective masses of proton and neutron provided by the Furusawa-Togashi EOS \citep{furusawaNewEquationState2017}.

\begin{figure*}
  \begin{minipage}{1\linewidth}
    \centering
    \includegraphics[width=1\textwidth]{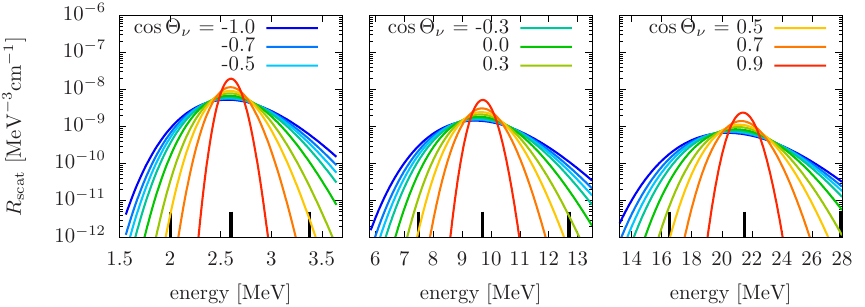}
    \subcaption{}
    \label{fig::reaction_rate_ene}
  \end{minipage}\\
  \begin{minipage}{1\linewidth}
    \centering
    \includegraphics[width=1\textwidth]{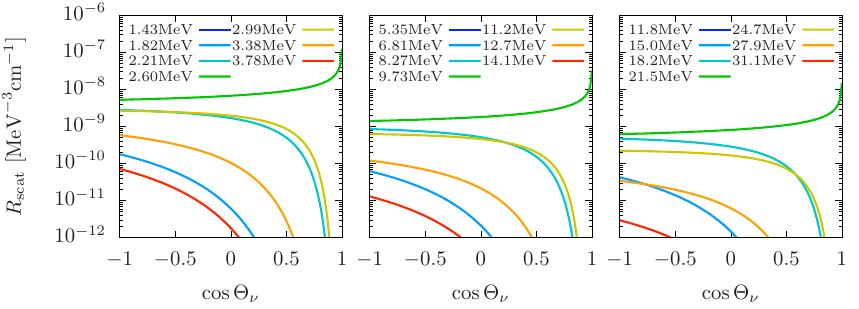}
    \subcaption{}
    \label{fig::reaction_rate_theta}
  \end{minipage}
  \caption{ The reaction kernel for the fixed incident energy:
  $\epsilon = 2.60 \, \mathrm{MeV}$, $\epsilon = 9.73 \, \mathrm{MeV}$ and $\epsilon = 21.5 \, \mathrm{MeV}$ in the left, middle and right columns, respectively.
  In the upper row,
  it is plotted as a function of the energy of the outgoing neutrino for some fixed scattering angles.
  The reddish (bluish) lines represent the forward (backward) scatterings.
  In the lower row, the kernel is plotted as a function of the scattering angle for fixed energies of the outgoing neutrino.
  The reddish (bluish) lines denote the up-scatterings (down-scatterings). }
  \label{fig::reaction_rate_all}
\end{figure*}

Figure~\ref{fig::reaction_rate_all} shows the reaction kernel $R_{\mathrm{scat}} (\epsilon,\epsilon^\prime,\cos\Theta_\nu)$, where $\Theta_{\nu}$ is the angle that the momentum of the outgoing neutrino makes with that of the incident neutrino, for different incident energies.
In the upper panels, we fix the scattering angle $\Theta_\nu$ in addition to the incident energy $\epsilon$ and plot the scattering kernel as a function of $\epsilon^\prime$, the energy of the outgoing neutrino, for different choices of $\Theta_{\nu}$.
The reddish and bluish lines represent the forward and backward scatterings, respectively.
It is apparent that the reaction rate drops quickly as the modulus of the energy transfer, $| \epsilon - \epsilon^\prime |$, increases, which indicates that the recoil of nucleon is indeed small.
It is still important for quantitatively accurate CCSN simulations as we explained earlier.
In fact, the width of the kernel in energy depends on the scattering angle rather sensitively and becomes larger for backward scatterings.
In the same panels, we show for reference the energy grid employed normally in our Boltzmann simulations.
Note that the incident energy is chosen to coincide with one of the cell interfaces of that grid.
The widths of the adjacent energy bins are much larger than that of the reaction kernel for the forward scatterings, $\cos \Theta_{\nu} >0$, in all cases. 
In the backward scatterings with $\cos \Theta_{\nu} \lesssim -0.5$, the kernel extends beyond the next bins, accounting for $\gtrsim 10 \% $ of these scatterings in some cases, which may not be negligible.

In the lower panels, we show the kernel as a function of the scattering angle for the fixed energies of incident and outgoing neutrinos.
It is again apparent that the scattering depends on the scattering angle.
For most of the values of $\epsilon^\prime$, the backward scatterings are dominant.
The exception is the scatterings with small energy transfer, which tend to occur in the forward direction.

\begin{figure} 
  \begin{minipage}{1\linewidth}
    \centering
    \includegraphics[width=1\textwidth]{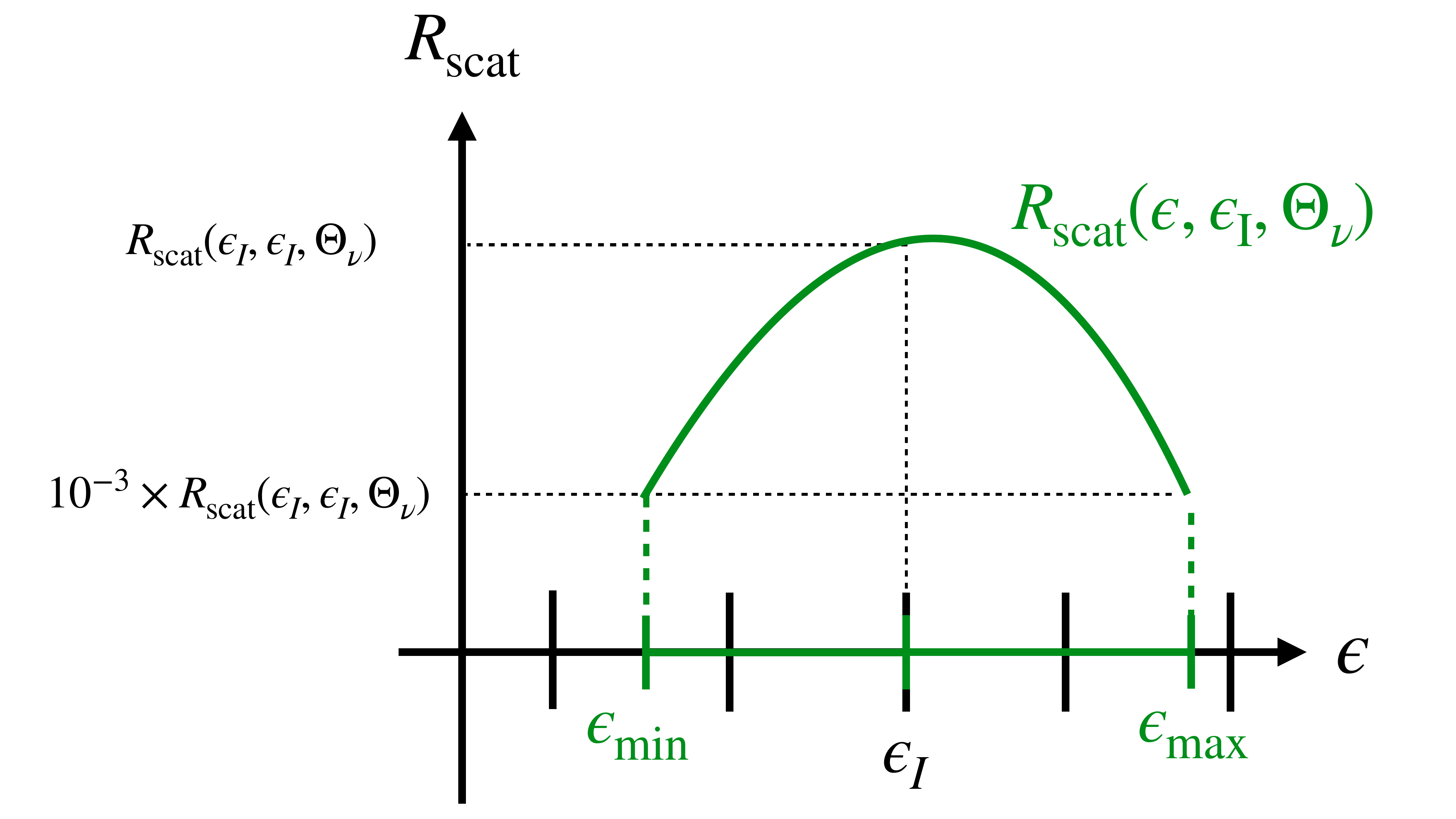}
    \subcaption{ Determination of the energy range.}
    \label{fig::subgrid_energy_range}
  \end{minipage}\\
  \begin{minipage}{1\linewidth}
    \centering
    \includegraphics[width=1\textwidth]{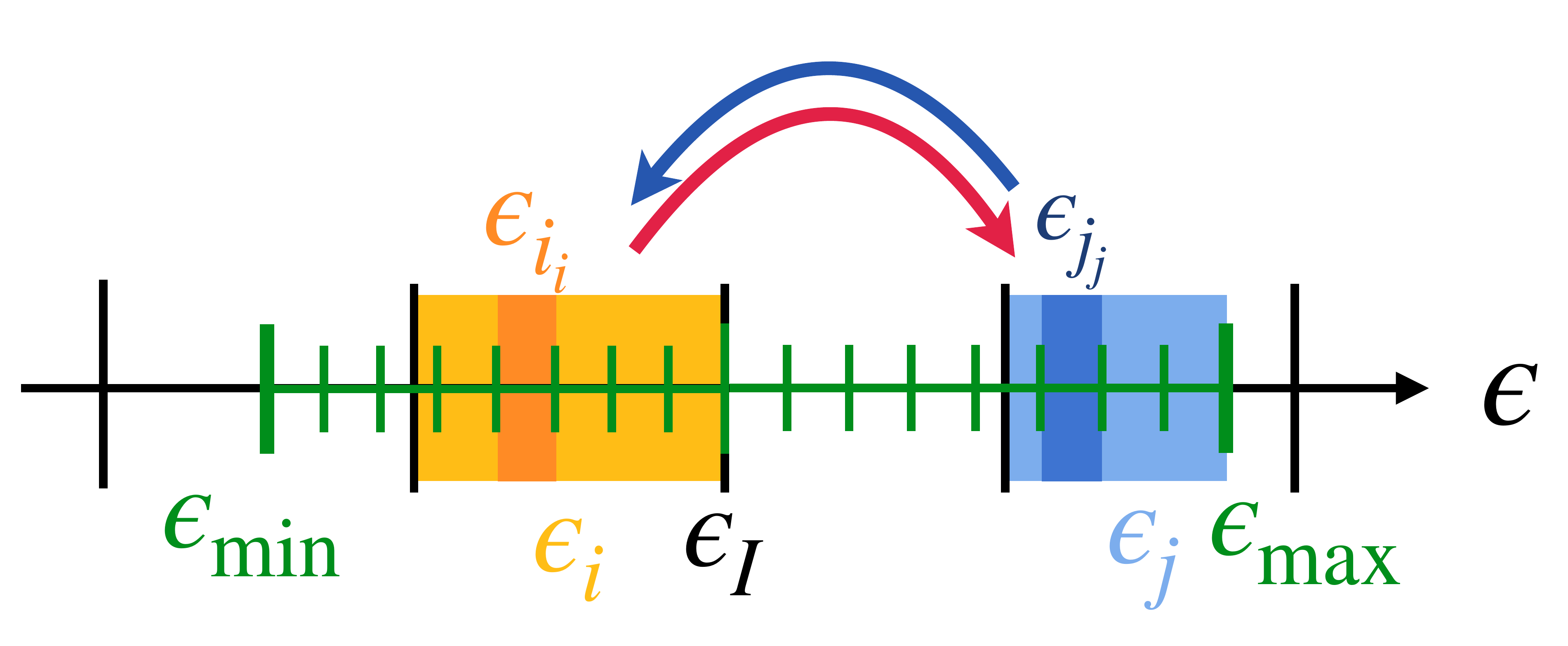}
    \subcaption{The subgrid model.}
    \label{fig::subgrid_model}
  \end{minipage}
  \caption{ (a) Schematic picture of the determination of the energy range spanned by the subgrid. For each energy \it interface \rm $\epsilon_I$, the values of $\epsilon_{\mathrm{min}}$ and $\epsilon_{\mathrm{max}}$ are determined. They also depend on the scattering angle.
  (b) Schematic picture of the calculations with the subgrid of down-scatterings (blue arrow) from the $j$-th cell to the $i$-th cell in the original grid and up-scatterings (red arrow) from the $i$-th cell to the $j$-th cell. The energy range $[\epsilon_{\mathrm{min}},\epsilon_{\mathrm{max}}]$ of the subgrid is determined as in (a). The subcells are created in the ranges $[\epsilon_{\mathrm{min}},\epsilon_I]$ and $[\epsilon_I,\epsilon_{\mathrm{max}}]$, individually. The summation with respect to $\epsilon_{i_i}$ in Eqs. \eqref{eq::reconst_reac_f} and \eqref{eq::reconst_reac_ffp} runs over the finer cells in the orange region whereas the summation with respect to $\epsilon_{j_j}$ runs over those in the blue region. The leftmost subcells in the orange and blue regions are overlapped with those regions only partially. Their contributions are reduced according to the overlaps.
  }
  \label{fig::subgrid_all}
\end{figure}

In our previous simulations, we treated this scattering as iso-energetic \citep{bruennStellarCoreCollapse1985}, since the energy transfer between neutrino and nucleon was thought to be smaller than the cell width of the energy grid with the current standard resolution.
In most of the recent simulations by other authors, the small energy transfer is treated approximately via the formula given by \cite{horowitzWeakMagnetismAntineutrinos2002}.
\cite{burasTwodimensionalHydrodynamicCorecollapse2006} were the first to introduce an energy-subgrid treatment in the truncated moment method and others followed suit
\citep{melsonNeutrinodrivenExplosion202015,Bollig2021,Janka2024,Janka2025}.
It seems that its accuracy has not been explored yet in the literature.
In this paper, we also adopt a subgrid method, which is similar to but is more involved than that for the truncated moment method, since the angular dependence is retained in our method.
At each \it interface \rm in the original energy grid we deploy a small subgrid with finer cells, the widths of which are determined from the width of the scattering kernel as depicted schematically in Fig.~\ref{fig::subgrid_all} and explained more in detail shortly.

At each \it interface \rm of the original energy grid and for all the scattering angles that the momenta of the incident and outgoing neutrinos on the angular grid points make, we first determine the energy range $ [ \epsilon_{\mathrm{min}},\epsilon_{\mathrm{max}} ]$ that the energy subgrid should cover.
The values of $\epsilon_{\mathrm{min}}$ and $\epsilon_{\mathrm{max}}$ are determined so that $R_{\mathrm{scat}} (\epsilon_{\mathrm{min}},\epsilon_I,\Theta_\nu)$ and $R_{\mathrm{scat}} (\epsilon_{\mathrm{max}},\epsilon_I,\Theta_\nu)$ should be equal to $10^{-3} \times R_{\mathrm{scat}} (\epsilon_I,\epsilon_I,\Theta_\nu)$ (see Fig.~\ref{fig::subgrid_energy_range}). 
We deploy the same number of energy bins in the energy range, $ [ \epsilon_{\mathrm{min}},\epsilon_I ]$ and $ [ \epsilon_I, \epsilon_{\mathrm{max}} ]$ individually, which is denoted by $N_{\mathrm{sub}}$.
Then we calculate for each scattering angle $\Theta_\nu$ the total number of down-scatterings across the cell \it interface \rm $\epsilon_{I}$, i.e., the scatterings from the energy range of $[\epsilon_I, \epsilon_{\mathrm{max}}]$ to that of $[\epsilon_{\mathrm{min}}, \epsilon_I]$ and the number of the upscatterings from $[\epsilon_{\mathrm{min}}, \epsilon_I]$ to $[\epsilon_I, \epsilon_{\mathrm{max}}]$ individually (see Fig.~\ref{fig::subgrid_model}).
Note that these energy ranges may cover more than one original cells as depicted in Fig.~\ref{fig::subgrid_energy_range}.
In such a case, we take all the latter cells into account (see Eqs. \eqref{eq::reconst_reac_f} and \eqref{eq::reconst_reac_ffp} below).

In so doing, we need the neutrino distribution.
We approximate it in each original energy bin with the Fermi-Dirac distribution modulo the overall normalization; its chemical potential is set to the value in $\beta$-equilibrium for $\nu_e$.
Note that we treat $\nu_e$ alone in this paper. 
The choice should be different for other species of neutrinos when they are incorporated.
The scaling of the distribution is done for the linear and nonlinear terms on the right hand side of Eq.~\eqref{eq:BoltzmannEquation} separately. The latter is originated from the Pauli-blocking factor. For the linear term the distribution function is scaled so that the number of neutrinos in the cell should be reproduced.
The scaling of the nonlinear term is determined so that the integral of the product of the distribution functions on the (fine) subgrid should coincide with the product of the actual distribution functions on the original (coarse) energy grid (see Eq.~\eqref{eq::reconst_reac_ffp}).
Note that only these scalings are done in real time during the simulation.

The collision terms on the original (coarse) energy grid may be now expressed as follows:
\begin{eqnarray}
&& C( f^{n+1} (\epsilon_i,\Omega_m) ) = \sum_{k=1}^{N_{\epsilon}} \frac{\Delta \displaystyle \frac{\epsilon_{k}^{3}}{3}}{(2 \pi )^2} \sum_{l=1}^{N_{\theta_\nu} \times N_{\phi_\nu} }  \Delta \Omega_l \nonumber \\
&& \times \left( R^{f}_{\mathrm{eff}}(\epsilon_k,\Omega_l; \epsilon_i,\Omega_m) f^{n+1} (\epsilon_k,\Omega_l) \right. \nonumber \\
&& - R^{ff}_{\mathrm{eff}}(\epsilon_k,\Omega_l; \epsilon_i,\Omega_m) f^{n+1} (\epsilon_k,\Omega_l) f^{n+1} (\epsilon_i,\Omega_m) \nonumber \\
&& - R^{f}_{\mathrm{eff}}(\epsilon_i,\Omega_m; \epsilon_k,\Omega_l) f^{n+1} (\epsilon_i,\Omega_m) \nonumber \\
&& \left. + R^{ff}_{\mathrm{eff}}(\epsilon_i,\Omega_m; \epsilon_k,\Omega_l) f^{n+1} (\epsilon_i,\Omega_m) f^{n+1} (\epsilon_k,\Omega_l) \right),
\end{eqnarray}
where the effective scattering kernels, $R^f_{\mathrm{eff}}$ and $R^{ff}_{\mathrm{eff}}$,  are given for $\epsilon_j \leq \epsilon_i$ as
\begin{equation}
R^f_{\mathrm{eff}} (\epsilon_i;\epsilon_j) = 
\displaystyle\frac{ \displaystyle \sum_{i_i,j_j} \displaystyle\frac{\Delta\epsilon_{i_i}^3}{(2\pi)^3} \frac{\Delta\epsilon_{j_j}^3}{(2\pi)^3} R_{\mathrm{scat}}(\epsilon_{i_i},\epsilon_{j_j}) f_{\mathrm{eq}}(\epsilon_{i_i})}{\displaystyle\frac{\Delta \epsilon_i^3}{(2\pi)^3} \frac{\Delta \epsilon_j^3}{(2\pi)^3} f_{\mathrm{eq}}(\epsilon_i) },
\label{eq::reconst_reac_f}
\end{equation}
\begin{equation}
R^{ff}_{\mathrm{eff}} (\epsilon_i;\epsilon_j) = 
\displaystyle\frac{ \displaystyle \sum_{i_i,j_j} \displaystyle\frac{\Delta\epsilon_{i_i}^3}{(2\pi)^3} \frac{\Delta\epsilon_{j_j}^3}{(2\pi)^3} R_{\mathrm{scat}}(\epsilon_{i_i},\epsilon_{j_j}) f_{\mathrm{eq}}(\epsilon_{i_i})f_{\mathrm{eq}}(\epsilon_{j_j})}{\displaystyle\frac{\Delta \epsilon_i^3}{(2\pi)^3} \frac{\Delta \epsilon_j^3}{(2\pi)^3} f_{\mathrm{eq}}(\epsilon_i) f_{\mathrm{eq}}(\epsilon_j) },
\label{eq::reconst_reac_ffp}
\end{equation}
respectively, whereas for $\epsilon_j > \epsilon_i$ they are given by the detailed balance condition as 
\begin{equation}
    R^f_{\mathrm{eff}} (\epsilon_i;\epsilon_j) = R^f_{\mathrm{eff}} (\epsilon_j;\epsilon_i) \exp \left( \frac{\epsilon_i - \epsilon_j}{T} \right),
\end{equation}
\begin{equation}
    R^{ff}_{\mathrm{eff}} (\epsilon_i;\epsilon_j) = R^{ff}_{\mathrm{eff}} (\epsilon_j;\epsilon_i) \exp \left( \frac{\epsilon_i - \epsilon_j}{T} \right).
    \label{eq::detailed_balance_reconst_reac_ffp}
\end{equation}

As mentioned earlier, the $j$-th energy cell may not be restricted to the cells next to the $i$-th cell, i.e., $j = i \pm 1$.
In Eqs.~\eqref{eq::reconst_reac_f}-\eqref{eq::detailed_balance_reconst_reac_ffp}, the angle arguments are omitted for notational simplicity; $f_{\mathrm{eq}}(\epsilon) = 1/ (1 + \exp{ (\frac{\epsilon - \mu}{T}}) ) $
is the Fermi-Dirac distribution function in $\beta$-equilibrium, where the neutrino chemical potential $\mu$ is given by the chemical potential of proton, $\mu_p$, electron, $\mu_e$ and neutron $\mu_n$ as $\mu = \mu_p + \mu_e - \mu_n$;
$\epsilon_{i_i}$ and $\epsilon_{j_j}$ are the energies at the subgrid cells inside the $i$-th and $j$-th cells of the original grid, respectively (see Fig.~\ref{fig::subgrid_model}).
Since the boundaries of the subgrids do not coincide with the interfaces of the original energy grid, some subcells may overlap with more than one original cells.
In that case the contributions from those subcells are distributed to the latter cells according to the overlapped ranges (see Fig.~\ref{fig::subgrid_model} again).

Now we determine how many cells to be deployed in the energy-subgrid.
For this purpose, we evaluate the relative error in the following numerical integration of the scattering kernel on the subgrid:
\begin{equation}
\sigma(\epsilon,\cos\Theta_\nu) = \frac{1}{(2\pi)^3} \int_{\epsilon_{\mathrm{min}}}^{\epsilon_{\mathrm{max}}} 2 \pi {\epsilon^\prime}^2 R_{\mathrm{scat}}(\epsilon,\epsilon^\prime,\cos\Theta_\nu) d\epsilon^\prime, \label{eq:cross_section}
\end{equation}
where the incident energy $\epsilon$ is chosen from the cell \it interface \rm values in the original energy grid.
We calculate this integral numerically for different combinations of $\epsilon$ and $\Theta_{\nu}$ by the piecewise quadrature.
Varying $N_{\mathrm{sub}}$, the number of cells in the energy-subgrid, we evaluate the relative error defined as
\begin{equation}
    \left| \frac{ \sigma_{\mathrm{num}}(\epsilon,\cos\Theta_\nu)- \sigma_{\mathrm{ex}}(\epsilon,\cos\Theta_\nu) }{\sigma_{\mathrm{ex}}(\epsilon,\cos\Theta_\nu)} \right|,
\end{equation}
where $\sigma_{\mathrm{num}}$ and $\sigma_{\mathrm{ex}}$ are the numerical and exact values, respectively;
the latter is obtained with a very large number of grid points ($N_{\mathrm{sub}} = 2^{17} \approx 10^5$).

\begin{figure}
  \begin{minipage}{1\linewidth}
    \centering
    \includegraphics[width=1\textwidth]{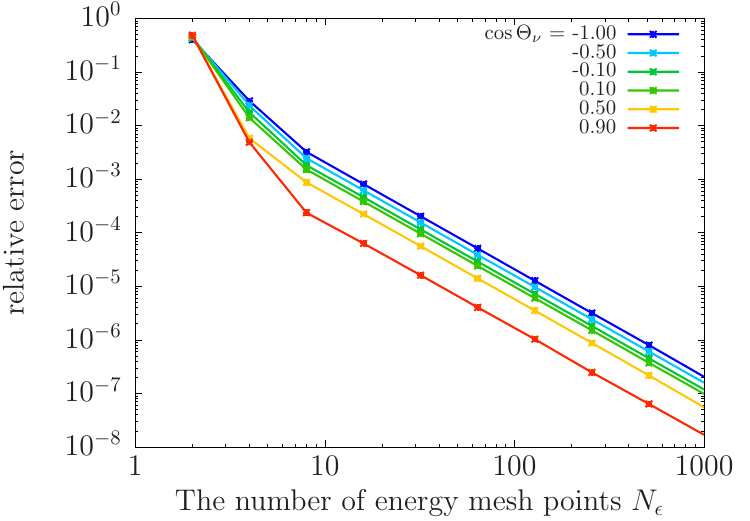}
    \subcaption{$\epsilon = 3.86 \mathrm{MeV}$}
    \label{fig::subgrid_relative_error_3.86MeV}
  \end{minipage}\\
  \begin{minipage}{1\linewidth}
    \centering
    \includegraphics[width=1\textwidth]{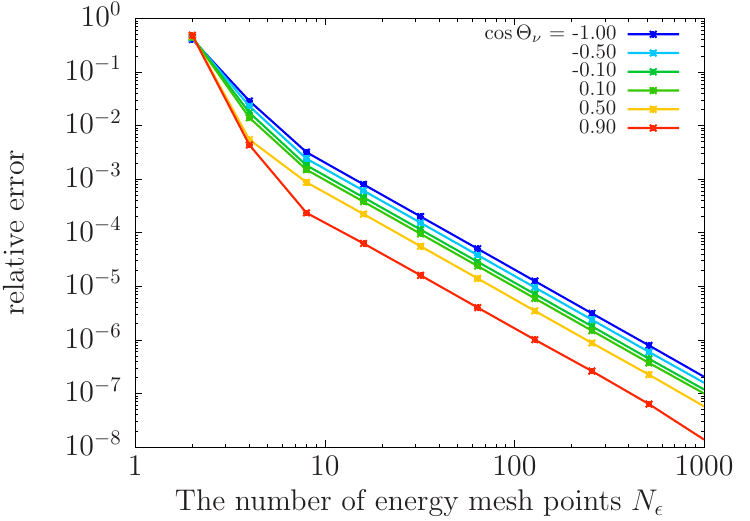}
    \subcaption{$\epsilon = 11.1 \mathrm{MeV}$}
    \label{fig::subgrid_relative_error_11.1MeV}
  \end{minipage}
  \begin{minipage}{1\linewidth}
    \centering
    \includegraphics[width=1\textwidth]{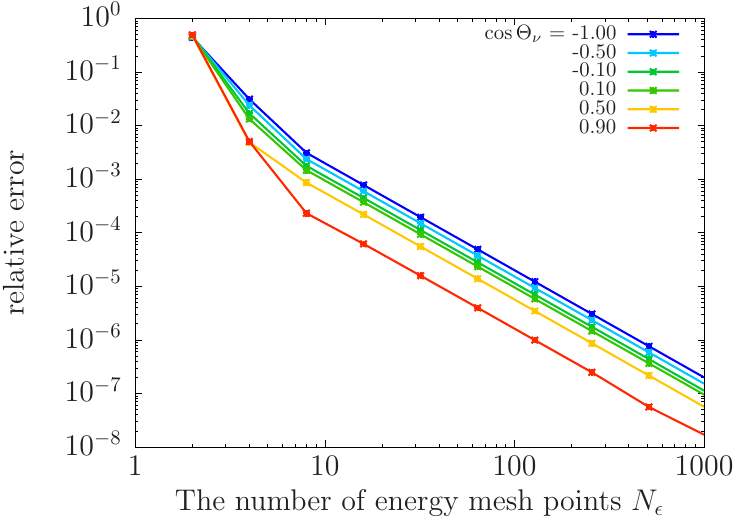}
    \subcaption{$\epsilon = 41.5 \mathrm{MeV}$}
    \label{fig::subgrid_relative_error_41.5MeV}
  \end{minipage}\\
  \caption{ The relative error of $\sigma$ in Eq.~\eqref{eq:cross_section} as functions of the number of energy mesh points for (a) $\epsilon = 3.86 \mathrm{MeV}$, (b) $11.1\mathrm{MeV}$ and (c) $41.5 \mathrm{MeV}$. The color coding is the same as in Fig.~\ref{fig::reaction_rate_all}. }
  \label{fig::subgrid_relative_error}
\end{figure}

The results are presented in Fig.~\ref{fig::subgrid_relative_error}.
One finds that the relative error decreases quadratically with $N_{\mathrm{sub}}$ irrespective of the values of $\epsilon$ and $\Theta_{\nu}$. 
The error is larger for backward scatterings in general.
This arises from the fact that the peak of the scattering kernel, which gives the largest errors in the piecewise quadrature, is more shifted from the cell interface to the cell center for the backward scatterings than for the forward scatterings.
We find at $ N_{ \mathrm{sub} } = 8 $ that the relative error gets $ \lesssim 5 \times 10^{-3} $, which is sufficiently small normally for the CCSN simulation.
We henceforth employ $ N_{\mathrm{sub}} = 8 $ in the energy-subgrid.
It is incidentally noted that in the previous implementations by other authors (e.g., \cite{burasTwodimensionalHydrodynamicCorecollapse2006}), the original mesh cells are subdivided into $ N_{ \mathrm{sub} } $ subcells, where $ N_{ \mathrm{sub} } $ is typically 6, which may not always be sufficient particularly for the forward scatterings, for which the kernel is narrow-peaked near the original cell-interface.

\end{document}